\documentclass[
 reprint,
 amsmath,
 pre,
 amssymb,
 nofootinbib,
 aps,
 longbibliography,
]{revtex4-1}

\usepackage[pdftex]{graphicx}
\usepackage{wasysym}
\usepackage{amsfonts}
\usepackage{ifsym}
\usepackage{pifont}
 \usepackage{footmisc}

\makeatletter
\newlength{\apb@width}
\newcommand{\autoparbox}[2][c]{\settowidth{\apb@width}{#2}\parbox[#1]{\apb@width}{#2}}

\makeatother

\newcommand{\namedref}[2]{\hyperref[#2]{#1~\ref*{#2}}}

\newcommand{\dd}{\mathrm{d}}

\newcommand{\Csphere}{{}^\bullet\kern-1.2pt C}
\newcommand{\Ctorus}{{}^\circ\kern-1.2pt C}

\newcommand{\nn}{\nonumber}

\newcommand{\COMMENT}[1]{}

\newcommand{\neqa}{\nonumber\end{eqnarray}}
\newcommand{\la}[1]{\label{#1}}

\newcommand{\<}{{\langle}}
\renewcommand{\>}{{\rangle}}

\newcommand{\re}{\relax{\rm I\kern-.18em R}}

\def\su2{{SU(2)}}

\def\[{\left[}
\def\]{\right]}

\def\({\left(}
\def\){\right)}
\def\[{\left[}
\def\]{\right]}

\def\<{\langle}
\def\>{\rangle}

\def\2F1{\,_2{\rm F}_1}

\usepackage[usenames,dvipsnames]{xcolor}
\usepackage{mathbbol}

\usepackage{cancel}
\usepackage{tikz}
\usepackage{ctable}
\usepackage{booktabs}
\usepackage[caption=false]{subfig}
\newcolumntype{L}[1]{>{\raggedright\let\newline\\\arraybackslash\hspace{0pt}}m{#1}}
\newcolumntype{C}[1]{>{\centering\let\newline\\\arraybackslash\hspace{0pt}}m{#1}}
\newcolumntype{R}[1]{>{\raggedleft\let\newline\\\arraybackslash\hspace{0pt}}m{#1}}

\newcommand{\beq}{\begin{equation}}
\newcommand{\eeq}{\end{equation}}
\newcommand{\beqq}{\begin{equation*}}
\newcommand{\eeqq}{\end{equation*}}
\newcommand\beqa{\begin{eqnarray}}
\newcommand\eeqa{\end{eqnarray}}
\newcommand\beqaa{\begin{eqnarray*}}
\newcommand\eeqaa{\end{eqnarray*}}
\newcommand\bea{\begin{array}}
\newcommand\eea{\end{array}}

\begin{document}


\title{Where is M-theory in the space of scattering amplitudes?}

\author{Andrea Guerrieri$^{a,b,c}$, 
Harish Murali$^{b,d}$, Jo\~ao Penedones$^{e}$, Pedro Vieira$^{b,f}$ 
}
\affiliation{$^{a}$ School of Physics and Astronomy, Tel Aviv University, Ramat Aviv 69978, Israel}
\affiliation{$^{b}$ Perimeter Institute for Theoretical Physics, 31 Caroline St N Waterloo, Ontario N2L 2Y5, Canada}
\affiliation{$^{c}$ Dipartimento di Fisica e Astronomia, Universita degli Studi di Padova, \& Istituto Nazionale di
Fisica Nucleare, Sezione di Padova, via Marzolo 8, 35131 Padova, Italy}
\affiliation{$^{d}$ Department of Physics and Astronomy, University of Waterloo, Waterloo, Ontario, N2L 3G1, Canada}
\affiliation{$^{e}$ Fields and String Laboratory, Institute of Physics, \'Ecole Polytechnique F\'ed\'erale de Lausanne (EPFL), \\
Rte de la Sorge, BSP 728, CH-1015 Lausanne, Switzerland}
\affiliation{$^{f}$Instituto de F\'isica Te\'orica, UNESP, ICTP South American Institute for Fundamental Research, Rua Dr Bento Teobaldo Ferraz 271, 01140-070, S\~ao Paulo, Brazil}


\begin{abstract}
We use the S-matrix bootstrap to carve out the space of unitary, {analytic}, crossing symmetric and supersymmetric graviton scattering amplitudes in nine, ten and eleven dimensions. 
We extend and improve the numerical methods of our previous work in ten dimensions. 
A key new tool employed here is  unitarity in the celestial sphere. 
In all dimensions, we find that the minimal allowed value of the
Wilson coefficient $\alpha$, controlling the leading correction to maximal supergravity, 
is very close but not equal to the minimal value realized in Superstring theory or M-theory.
This small difference may be related to inelastic effects that are not well described by our numerical extremal amplitudes.
Although $\alpha$ has a unique value in M-theory, we found no evidence of an upper bound on $\alpha$ in 11D. 

\end{abstract}

\pacs{Valid PACS appear here}
\maketitle

\section{Introduction} \label{sec:introduction}

General relativity (GR) is a low energy effective field theory.
What are its  consistent UV completions?
In \cite{ourST}, we addressed this question in 10D and with maximal supersymmetry, by studying the allowed space of the leading Wilson coefficient $\alpha$ that controls the first corrections to supergravity.
In this paper, we extend this approach to other dimensions, namely 9D and 11D.
In principle, we can study any $5\le d \le 11$ but for $d\le 8$ the 1-loop corrections appear at the same or lower order than $\alpha$ which makes the numerics more challenging. 

Let us briefly review the setup explained in  \cite{ourST}.
The two-to-two scattering amplitude of the graviton multiplet for any $d\ge 5$ theory with maximal SUSY takes the form
\beq
\mathbb{A}_{2\to 2}= \textbf{R}^4 A(s,t,u) \,, \la{Adef}
\eeq
where  $\textbf{R}^4$ is a universal prefactor that takes care of the various components in the graviton multiplet.
It is convenient to consider the following scalar amplitude
\beq
\frac{T(s,t,u)}{8\pi G_N} 
=s^4\(\frac{1}{stu}+\alpha \,\ell_P^6+ \dots \) \equiv s^4 A(s,t,u) \label{defAlpha}
\eeq
where  
$16\pi G_N=(2\pi)^{d-3} \ell_P^{d-2}$.
The $\dots$  include the one loop contribution (see appendix \ref{unitAppendix}) and higher order Wilson coefficients (Wcs).


\section{String Theory}
\label{alphaST}

In this section, we review the String Theory predictions for $\alpha$ in $d=9,10,11$.

\noindent\rule[0.5ex]{\linewidth}{1pt}

\noindent $\mathbf{d=11}.$ In M-theory, the parameter $\alpha$ is fixed to
\beq
\label{alphaMtheory}
\alpha = \frac{(2\pi)^2}{3 \ 2^7} \simeq 0.1028\,.
\eeq
Here we follow \cite{Green:1997di}. 
It differs by the $2\pi$ numerator from \cite{Alday:2020tgi} which uses a slightly different convention for $\ell_P$. 

 \noindent\rule[0.5ex]{\linewidth}{1pt}
 
\noindent $\mathbf{d=10}.$ In type IIB superstring, $\alpha$ can be cast in terms of a non-holomorphic Eisenstein series \cite{Green:1997tv, Chester:2019jas}
\beq
\label{alphaIIB}
\alpha^\text{IIB}= \frac{1}{2^6} E_\frac{3}{2}(\tau,\bar{\tau}) = \frac{1}{2^6}  \sum\limits_{m,n \in \mathbb{Z} \atop (m,n)\neq (0,0)} \frac{\left({\rm Im}\, \tau\right)^\frac{3}{2}}{|m\tau+n|^3}
\eeq
where $\tau$ is the complexified string coupling, which can be taken in the fundamental domain~$\left| {\rm Re}\, \tau \right| \le 1$ and $|\tau|\ge 1$. 
The minimal value is attained at the ``corners" $\tau=e^{i\pi/3}$ and  $\tau=e^{2i\pi/3}$ where~\cite{AldayBissi} 
\beq
E_\frac{3}{2}(e^{i\pi/3},e^{-i\pi/3})=
{ 3^{\frac{1}{4}} \zeta
    (\tfrac{3}{2 })
   \left[\zeta
   \left(\tfrac{3}{2},\tfrac{1}{3}\right
   )-\zeta
   \left(\tfrac{3}{2},\tfrac{2}{3}\right
   )\right]}/{ \sqrt{2}}\,. \nn
\eeq
In type IIA superstring theory, we have (see e.g. \cite{Binder:2019mpb})
\beq
\label{alphaIIA}
\alpha^\text{IIA}= \frac{\zeta(3)}{32 g_s^{3/2}} + g_s^{1/2} \frac{\pi^2}{96}.
\eeq
This function attains its minimum value  $\frac{\pi^{3/2} (\zeta(3))^{1/4}}{24\sqrt{3}}$
for $g_s^2 = 9\zeta(3)/\pi^2$. Numerically, the minimum values of $\alpha$ in type IIB and type IIA are approximately $0.1389$ and $0.1403$ respectively.

\noindent\rule[0.5ex]{\linewidth}{1pt}

\noindent $\mathbf{d=9}.$ Here, we have \cite{Green:2010wi, Pioline:2015yea}
\beq
\alpha =\frac{1}{2^6}\left[ \nu^{-\frac{3}{7}} E_\frac{3}{2}(\tau,\bar{\tau}) +  \frac{2\pi^2}{3}\nu^\frac{4}{7}\right]
\label{alpha9D}
  \eeq
where $\nu$ is related to the compactification radius from 10 to 9 dimensions (see appendix \ref{reductions}).
It is easy to minimize \eqref{alpha9D} using the change of variable $\nu \to x^7 E_\frac{3}{2}(\tau,\bar{\tau})$ which leads to 
\beq
\alpha =\frac{\left[E_\frac{3}{2}(\tau,\bar{\tau})\right]^\frac{4}{7}}{2^6}\left[ x^{-3}   +  \frac{2\pi^2}{3} x^4\right] \ge 0.2417 
  \eeq

\noindent\rule[0.5ex]{\linewidth}{1pt}

These formulas   satisfy non-trivial (de)-compactifying relations as reviewed in appendix \ref{reductions}.

\section{Unitarity in the sky}
Our crossing symmetric ansatz  is the same as in~\cite{ourST}: 
 \beq
\frac{T(s,t,u)}{8 \pi G_N}{=} s^4 \Big( \underbrace{\frac{1}{s t u}}_{\texttt{SUGRA}}{+} \underbrace{ \prod_{A=s,t,u}\!\!(\rho_A{+}1)^2\!\!\! \!\!\sum_{a{+}b{+}c{\le} N}^{\prime} \alpha_{(abc)}\rho_s^a \rho_t^b \rho_u^c }_{\texttt{UV completion}}\Big) 
\label{ansatz0}
\eeq
where $\rho_z$ maps the complex $z$ plane minus the $z>0$ cut to the unit disk and where $\alpha_{(abc)}$ are our (symmetric) variables. 
If $N \to \infty$ this ansatz should cover \textit{any} amplitude \cite{4dpaper1}. The challenge is to make sure it is a good approximation for the amplitude when $N$ is finite. The ansatz does some wonderful things: it can satisfy unitarity for the first hundreds of spins and has a good large energy behavior at fixed angle as explained in \cite{ourST}. It  struggles with large energy and small angle or equivalently for a double scaling limit of large spin and large energy if we keep $N$ finite, we elaborate on this further in Appendix \ref{runawayappendix}. 

Unitarity comes from the condition 
\beq
\mathbb{I} = \mathbb{S}^\dagger \mathbb{S}  \,,
\eeq
for the full quantum gravity S-matrix $\mathbb{S} = \mathbb{I}+i \mathbb{T}$. Once we insert a complete basis of states we transform this equality into an infinite sum of matrix elements. If we truncate to $2\to 2$ particle amplitudes $\mathbb{t}\leftarrow \mathbb{T}$ we get 
\beq
2\,\text{Im} \,\mathbb{t} - \mathbb{t}^\dagger \mathbb{t} \succeq 0
\eeq
where the last inequality is understood as stating that the matrix on the left hand side is positive semi-definite. There are two obvious things we can do here. 

The first one (which has been used in all the recent S-matrix bootstrap explorations) is to diagonalize the left hand side by going to a basis of two particle states of definite spin. Defining $S_\ell(s)$ to be equal to~\footnote{$\mathcal{N}=2^{3-2 d}s^{\frac{d}2-2}\pi^{1-\frac{d}{2}}/\Gamma(\tfrac{d}{2}-1)$,  $C_\ell^\alpha$ are Gegenbauer's polynomials.}
\beq
\! 1+ i  \mathcal{N}\!\! \int\limits_{-1}^1\! \dd z (1-z^2)^{\frac{d-4}2} \frac{C_\ell^{\frac{d-3}2}(z)}{C_\ell^{\frac{d-3}2}(1)} T\big(s,s\tfrac{-z-1}{2},s\tfrac{-z+1}{2}\big)\,,  \la{SlDef} 
\eeq
we translate unitarity into the very simple condition 
\beq
|S_\ell(s)| \le 1 \label{Sunit} \,.
\eeq
We impose this condition for $\ell=0,2,4,\dots, L$ where $L$ is big. In practice we go as far as $L=200$.\footnote{One important bottle-neck is evaluating the integrals in (\ref{SlDef}); we do them monomial $\rho_s^a \rho_t^b \rho_u^c$ by monomial and save these integrals so we only need to do them once. Still, there are many monomials, many spins   and   these integrals need to be done for each $s$  with a huge precision to be fed into \texttt{sdpb} later. For illustration, our integral files for $L=200$, $N=30$ and $326$ grid values of $s$ occupy around $10$ gigabytes.}

The second thing we can do which is novel and proves very convenient is what we call {\color{blue}\textit{Unitarity in the Sky}}. It follows from noting that $2\,\text{Im}\, \mathbb{t} - \mathbb{t}^\dagger\mathbb{t}$ defines a positive operator no matter what in-going and out-going states we sandwich it with. Since the $\mathbb{t}^\dagger\mathbb{t}$ term is quadratic in the ansatz we obtain semidefinite constraints of size $M^2$ ($M$ being the number of terms in (\ref{ansatz0}) which scales as $N^2$ when restricted to on-shell external particles \cite{4dpaper1}) which quickly gets prohibitively expensive.

\begin{figure}[t]
    \centering
    \includegraphics[scale=0.12]{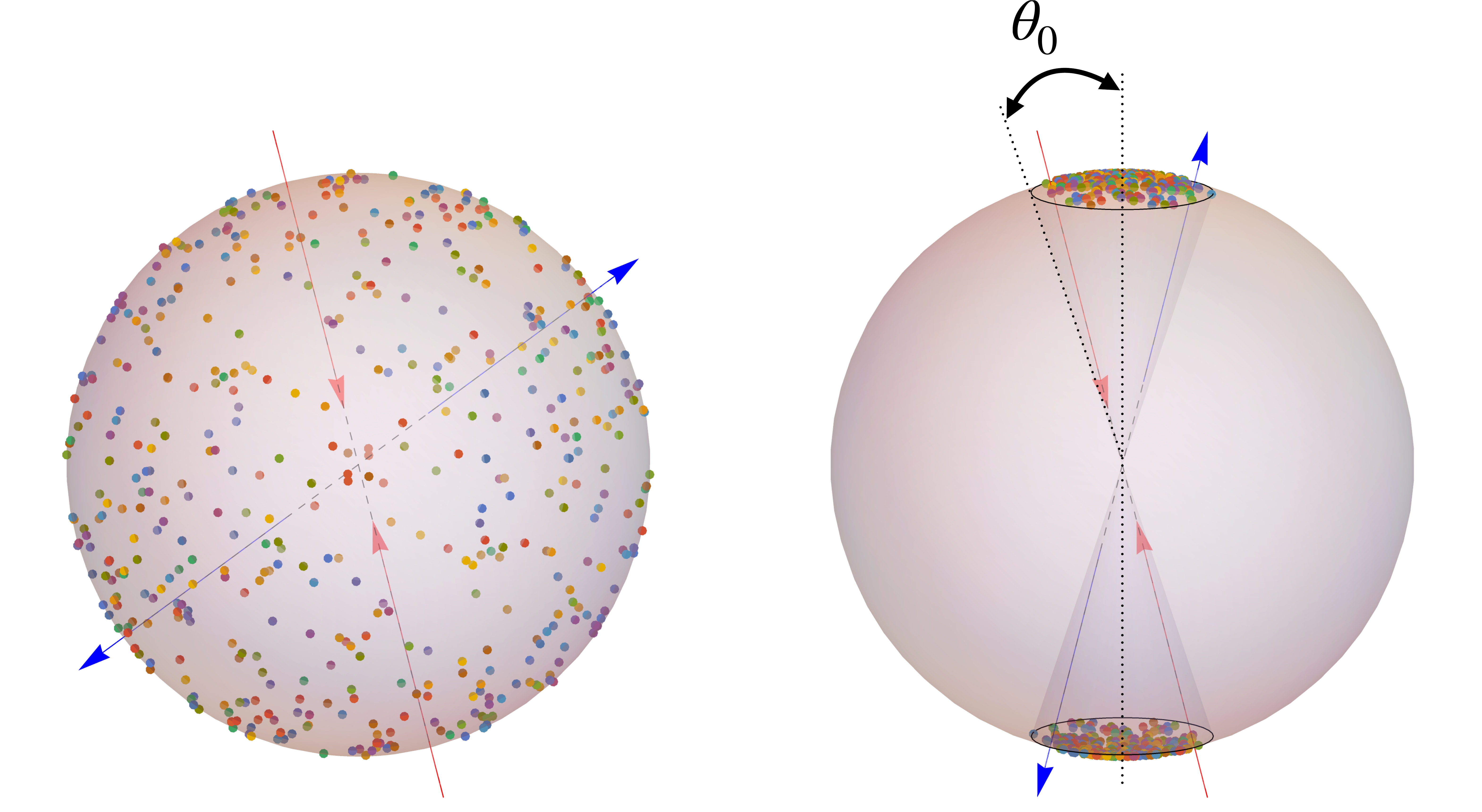}
    \caption{Distribution of momenta for \textit{Positivity in the Sky}. On the left, the momenta are uniformly chosen over the entire celestial sphere at infinity while on the right, the ingoing and outgoing momenta are chosen to lie within a cone of small opening angle $\theta_0$.}
    \label{fig:positivity}
\end{figure}

\begin{figure}[t]
    \centering
    \includegraphics[scale=0.185]{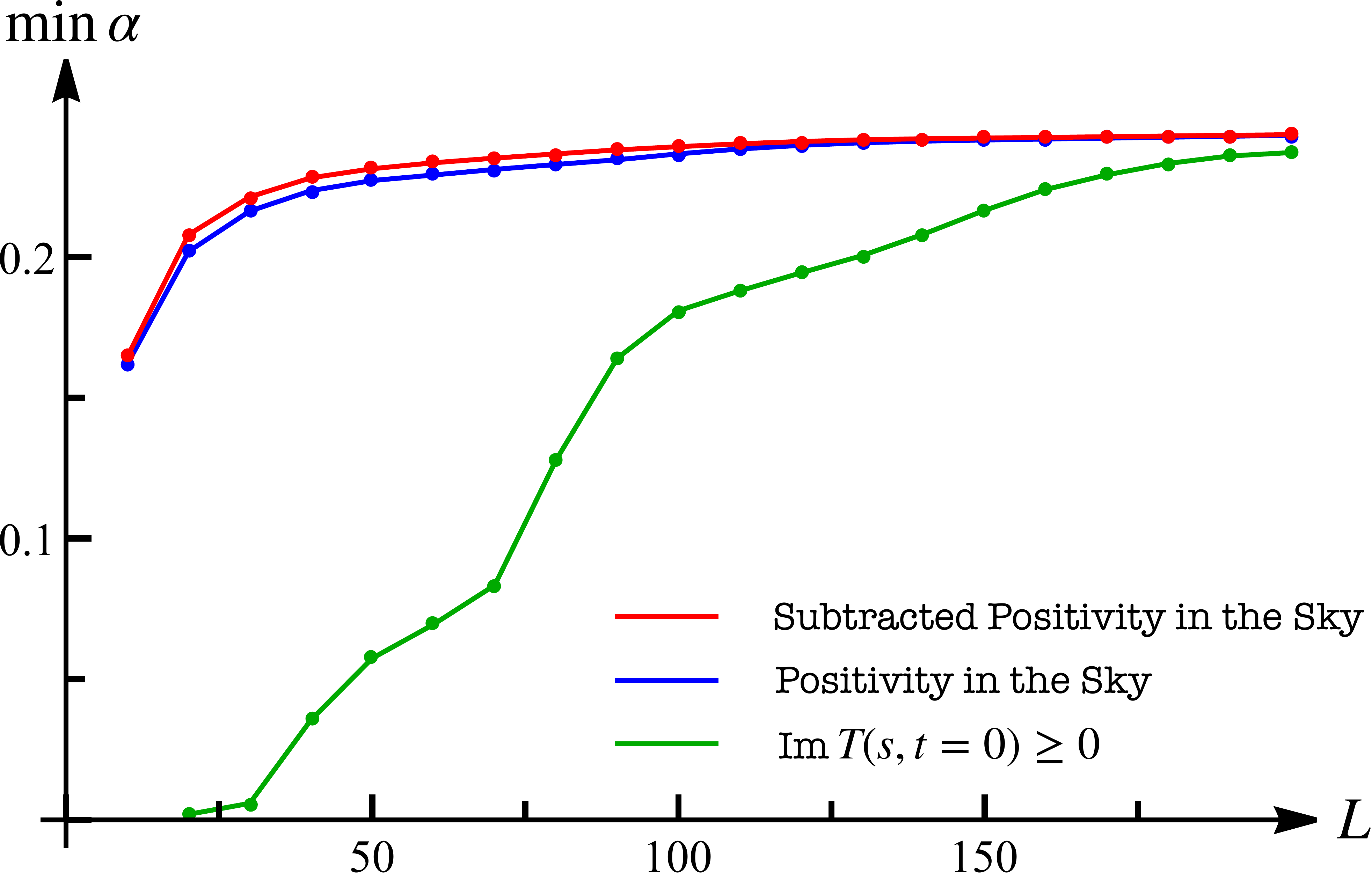}
    \caption{Minimum value of $\alpha$ as we impose more and more unitarity constraints up to spin $L$ for $N=24$ in $d=10$. The green curve from \cite{ourST} converges way slower than the red and blue curves where we supplement the physical conditions with the new \textit{Positivity in the Sky}. We see that \textit{Spin Subtracted} is slightly better but not significantly. This improved convergence allows us to restrict our numerics to $L\le 200$ while in \cite{ourST} one needed $L=300$ to attain reasonable convergence.
}
    \label{improvement}
\end{figure}

One workaround is to impose the weaker positivity condition $\text{Im}\, \mathbb{t}\succeq 0$. 
For each energy $s$ we consider $d$ random momenta\footnote{Less points would not probe the full angular unitarity in $d$ dimensions. More points would lead to heavier numerics.} in the celestial sphere peaked around the forward limit, see figure \ref{fig:positivity}. We then construct a $d\times d$ matrix $M$ with matrix elements $i,j$ equal to $\text{Im}(T)$ at energy $s$ and scattering angle $\theta_{ij}$ being the angle in the celestial sphere between vectors $i$ and $j$, 
\beq
\text{Im}(T)_{ij}=\text{Im}(T)\left(s,-s\tfrac{1-\cos\theta_{ij}}{2},-s\tfrac{1+\cos\theta_{ij}}{2}\right).
\eeq
The diagonal elements of the matrix being positive is nothing but the familiar optical theorem at forward scattering. We can now impose that all these random matrices, one for each grid point $s$, are positive semi-definite. These extra conditions which we dub as {\color{blue}\textit{Positivity in the Sky}} are very easy to impose because we do not need to compute tedious integrals and are numerically tractable because they grow linearly with the ansatz size $M$. These conditions capture  all spin information -- specially large spin since we focus close to the forward limit -- and thus greatly help with convergence in $L$ of the bootstrap as displayed in figure \ref{improvement}. 

A stronger condition we imposed is $\text{Im}\, \mathbb{t}-\text{Im} \mathbb{t}_L\succeq 0$ where $\mathbb{t}_L$ is the contribution of the first $L$ spins, for which we imposed full unitarity; in practice this means subtracting the sum over the first $L$ spins from the matrix $\text{Im}(T)_{ij}$ and imposing positivity of this subtracted matrix. This is not extra costly since these first $L$ integrals were computed to impose full unitarity \eqref{Sunit} anyway. We call this stronger yet rigorous condition {\color{blue}\textit{Spin Subtracted Positivity in the Sky}}.

We can also take extreme kinematical limits to get more mileage out of the unitarity constraints. As explained in \cite{ourST} an interesting limit of (\ref{SlDef}) which immediately yields several linear constraints on the $\alpha_{(abc)}$'s is to consider large energy $s$ at fixed spin $\ell$ or large spin $\ell$ at fixed energy $s$.\footnote{Note that the regime of large spin $\ell$ and fixed energy is rather tame - the partial waves here decay as $\ell^{2-d}$ (for massless scattering), and therefore imposing positivity of   the imaginary part is sufficient to ensure unitarity.} We imposed both in this paper as well. Importantly, they hold at finite~$N$.

We could also take large spin \textit{and} large energy. If we scale $s = x \ell^2$ and send $\ell \to \infty$ and expand the unitarity condition (\ref{Sunit}) perturbatively in $1/\ell$ we obtain a polynomial positivity condition. These conditions, valid for large $\ell$, can be recast as an SDP problem \cite{Poland:2011ey}. However, these conditions turn out to be too constraining at finite $N$, because they demand an infinite tower of resonances, analogous to the double twist operators in the Conformal bootstrap.  We discuss further this important point -- raised to us by Simon Caron-Huot who we thank -- and its implications for our numerics in appendix \ref{runawayappendix}.

\section{Results} 
\begin{figure*}
   \includegraphics[scale=0.405]{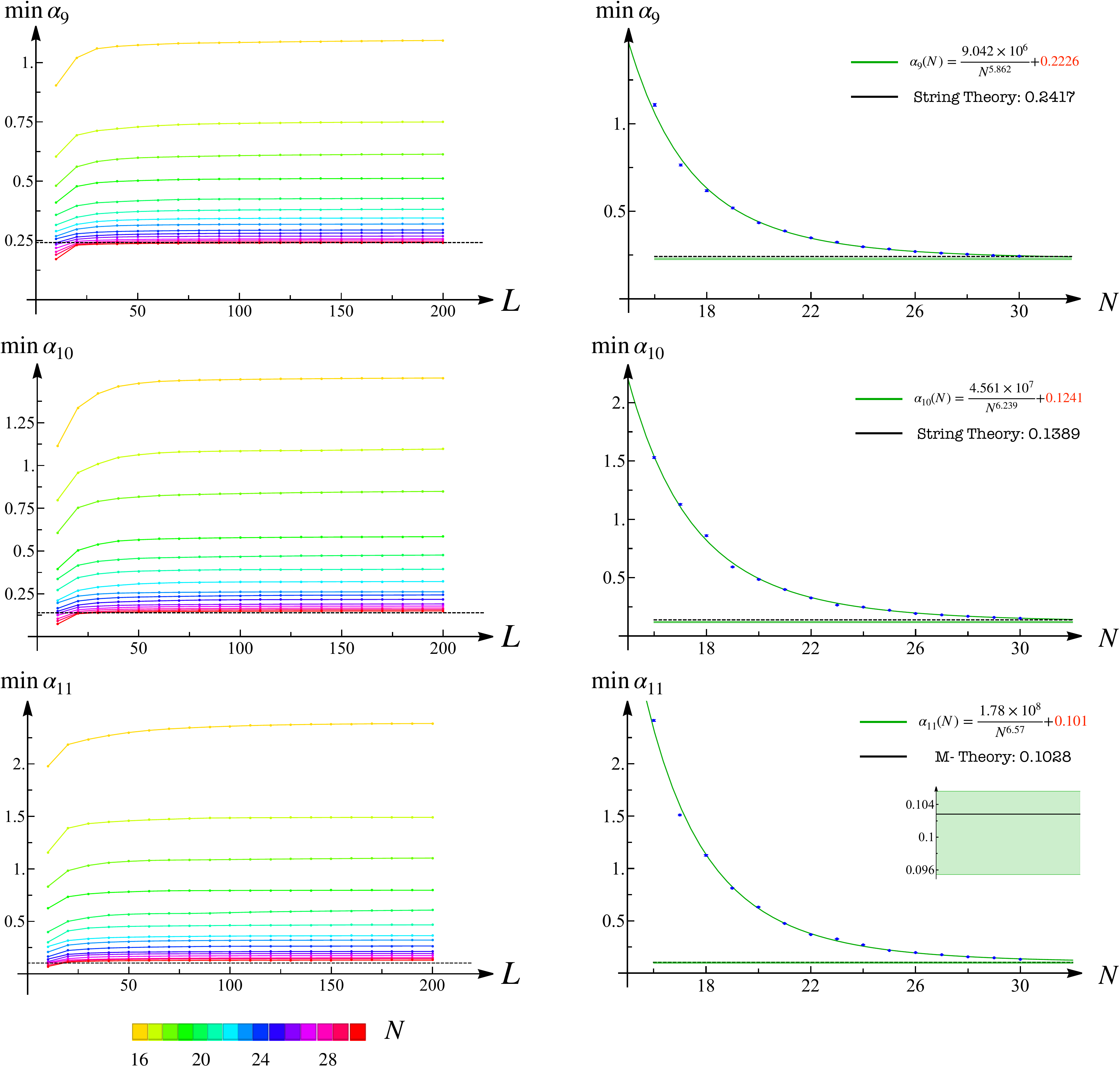}
   \caption{On the left, we have the data from minimizing $\alpha$ in 9, 10 and 11 dimensions plotted against $L$. The values of $N$ corresponding to the different colours are shown in the legend and the horizontal black lines are $\alpha_{\min}$ from string theory. The points on the right are obtained after fitting in $L$ and the error bars represent the variance across several fits. The large $N$ extrapolations of $\alpha_{\min}$ fall in the green (very thin) strips. Notably, in 11 dimensions the string theory result lies within the error bars of the bootstrap result, as shown in the bottom right inset} \label{ResultsAllD}
\end{figure*}

\la{results}
We minimize $\alpha$ subject to unitarity and positivity constraints using \texttt{sdpb} \cite{Landry:2019qug,Simmons-Duffin:2015qma}. Details about our numerical setup are found in Appendix \ref{ap:numerics}. The results for dimensions $d=9,10,11$ are summarized in figure \ref{ResultsAllD}. Each colour corresponds to an $N$ in the ansatz ranging from $16$ to $30$.

What these plots show are the numerics converging to some asymptotic value as we push $L$ further and further.\footnote{Strictly speaking these are transient plateaus as argued in the appendix \ref{runawayappendix} but that does not invalidate our conclusions as explained there.} The reader might have noticed that the numerics in figure \ref{ResultsAllD} converge way better than those in \cite{ourST} as far as $L$ goes. This is because of the new \textit{Spin Subtracted Positivity in the Sky} conditions we are imposing. Figure \ref{improvement} highlights this huge improvement for an illustrative $N=24$.

Having extrapolated in $L$ we extrapolate in $N$, see figure \ref{ResultsAllD}. These fits (first in $L$, then in $N$) are of course a bit of an art. We attach a notebook with these fits so that other artists can try it themselves. What we did was the same as in \cite{ourST}: we try a bunch of fits in $L$ and weight all of them by how well they fit the data. The dispersion gives us an estimate for the plateau as well as some error bars. Then we fit these points with their corresponding error bars to an ansatz of the form $A+B/N^C$ to extract the infinite $N$ estimate for $A$ which is our final prediction for the minimum value of $\alpha$ as extracted through this primal bootstrap approach. These fits are plotted in figure \ref{ResultsAllD}, right column, and the resulting bounds are summarized in the following table:
\begin{table}[h!]
\centering
\begin{tabular}{||c c c||} 
 \hline
Dimension & Bootstrap & String/M--Theory \\ [0.5ex] 
 \hline 
 \hline
 9 & $0.223\pm0.002$ & 0.241752\\
 \hline
 10 & $0.124\pm0.003$& 0.138949 \\
 \hline
 11 & $0.101\pm0.005$ & 0.102808\\
 \hline
\end{tabular}
\label{alphaResultTable}
\caption{The bootstrap minimal value estimates for $\alpha$ are very close to the minima attained by String/M-theory. They are slightly smaller; that small gap between the two might be due to unaccounted inelasticity effects as discussed below.}
\end{table}

Beautifully, the bounds are quite close to the String Theory predictions in all dimensions and of course they allow M-theory and String Theory at all couplings. They better! In fact the fit predicts a value slightly below the minimum String/M-theory value. We comment on that small gap in the next section when we analyse the extremal amplitudes in more detail.  

Although $\alpha$ in M-theory takes a precise value, we do not find any evidence for an upper bound. It should not come as a surprise: we are not incorporating all consequences of maximal supersymmetry in the Bootstrap. Instead, we could fix $\alpha$ in M-theory as an input, and explore the bounds on the higher derivative corrections.

\section{Extremal Amplitudes}
Once we minimize $\alpha$ we get much more than this minimum value as the outcome of the extremization: we get the full set of variables $\alpha_{(abc)}$ and thus the full amplitude. 

Instead of extrapolating $N$ and $L$ to infinity, in this section we will analyse the outcomes of the largest values of $N=30$ and $L=200$ in our numerics. If we take $\alpha$ as a good proxy for how close these are to the optimal large $N,L$ solutions, we would guess that taking these values already gives us a reasonable qualitative picture:
\begin{table}[h!]
\centering
\begin{tabular}{||c c c||} 
 \hline
Dimension & $(N,L)=(30,200)$ & Bootstrap Extrapolation  \\ [0.5ex] 
 \hline 
 \hline
 9 & 0.2411 & $0.223\pm0.002$ \\
 \hline
 10 & 0.1499 & $0.124\pm0.003$ \\
 \hline
 11 & 0.1304 & $0.101\pm0.005$ \\
 \hline
\end{tabular}
\label{alphaResultTable}
\caption{For $N=30$ and $L=200$ we are already about $80\%$ close to the optimal primal solution if we trust using $\alpha$ as a proxy of that convergence.}
\end{table}

\subsection{Resonances}
\begin{figure}[t]
    \centering
    \includegraphics[width=\linewidth]{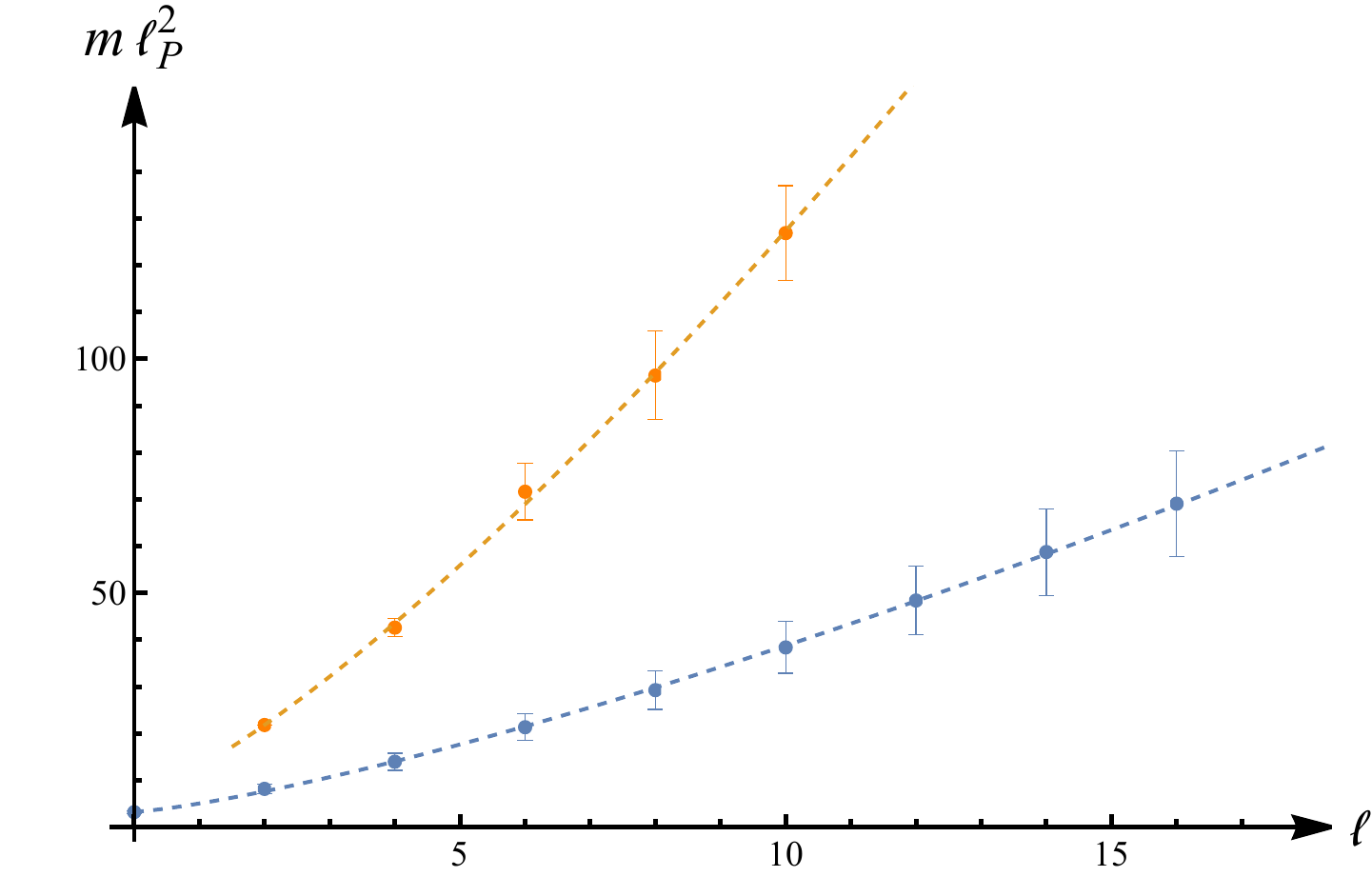}
    \caption{The first two Regge trajectories in 10d for N=30 and L=200. The error bars represent the widths and the resonances lie on curved trajectories that scale approximately like $\ell^{1.3}$. More details in appendix \ref{ap:resonances}. }
    \label{ReggeT}
\end{figure}
\begin{figure}[t]
\centering
        \includegraphics[scale=0.35]{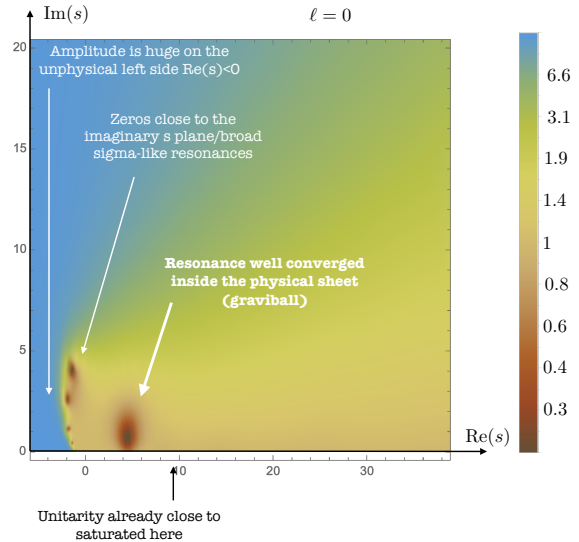}
\caption{$|S_0(s)|$ in the complex $s-$plane. For physical energies~$s>0$ unitarity tends to saturation with $|S_0(s)|\to 1$ as~$N\to \infty$. Here we are plotting $N=30$. We see smooth unitarity saturation with a graviball well inside the physical sheet with $m^2 \simeq 3+i$ in units of Planck mass. ($d=10$ here)}
\label{spin0}
\end{figure}
\begin{figure}[t]
\centering
        \includegraphics[scale=0.35]{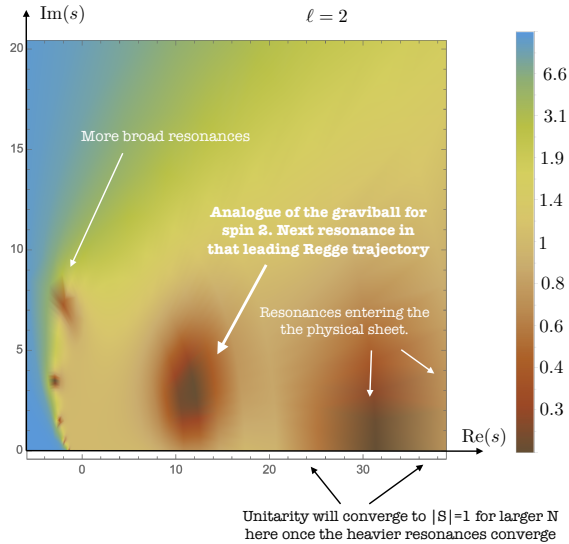}
\caption{The spin $2$ resonance in the Regge trajectory of the graviball is heavier and broader but still very well defined. There are now infinitely many resonances showing up. Unitarity is not as clean here close to heavier resonances for this same~$N=30$. Indeed, when resonances enter the physical sheet through the $s>0$ cut we must have $|S|=0$ at its location and thus sacrifice unitarity saturation momentarily. More details can be found in Appendix \ref{ap:resonances}.
($d=10$ here)}
\label{spin2}
\end{figure}

In figures \ref{spin0} and \ref{spin2} we depict the absolute value of the partial waves $S_{l}(s)$ with spin $0$ and $2$ respectively for $d=10$ in the complex $s$ plane. A more comprehensive set of such plots can be found in appendix \ref{ap:resonances}. A few qualitative features emerge:
\begin{itemize}
    \item Spin $0$ is special. It has a single well defined resonance. 
    \item There are infinitely many massive resonances at higher spins. 
    \item There are also broader resonances in the complex plane whose particle interpretation is less obvious. 
\end{itemize}
Let us now expand on these points. 

For $l=0$ the well defined resonance (i.e. zero of $|S_0|$) close to the real line is sometimes called the graviball \cite{Blas:2020dyg}. We find in the numerics in any dimension: \footnote{It is amusing to note that our lower bound $\alpha \geq 0.12$ in ten dimensions is not so far from the upper bound (3.42) in \cite{Caron-Huot:2021rmr} $\alpha\leq 0.09$ estimated using the mass of the graviball $m^2=3.2/\ell_P^2$ in Table III. At this point, it seems like a numerical coincidence given that the extremal amplitudes studied here are quite far from being weakly coupled. }
\begin{table}[h!]
\centering
\begin{tabular}{||c c||} 
 \hline
Dimension & $m^2 \ell_P^2$  \\ [0.5ex] 
 \hline 
 \hline
  9 & $3.0 + 0.8 i$ \\
 \hline
  10 & $3.2 + 0.5 i$ \\
 \hline
 11 & $3.3 + 0.3 i$ \\
 \hline
\end{tabular}
\label{GraviballTable}
\caption{Spin-0 ``graviball" in different dimensions. }
\end{table}\\
The $d=10$ value was estimated first in \cite{ourST}.\footnote{The real part is the same as in \cite{ourST} but the width here is better estimated; it is broader than originally reported.}

At higher energy for $l \geq 2$ there are more such zeros of~$|S_l(s)|$. These correspond to heavier resonances. There are probably infinitely many such resonances for any spin~$l$ but we will only really attain this as $N,L\to \infty$; in the current outcome of the numerics we can only resolve a few of these. As pointed out in \cite{ourST} these resonances seem to organize themselves in nice curved Regge trajectories, see figure \ref{ReggeT}.

That we find an infinite sequence of resonances for higher spins is perhaps not surprising. We know gravity requires them to Reggeize properly. It is amuzing though that spin $0$ seems to have a single resonance. Since we are minimizing $\alpha$ which can be extracted from a simple sum rule \cite{ourST}
\beq
 \alpha\, \ell_P^{6} =  \frac{2}{\pi} \int_{0}^\infty ds\, \frac{ {\rm Im}\, A(s+i\epsilon,t=0) }{s} \,,
\label{alphasumrule0}
\eeq
we see that minimizing the imaginary part of the amplitude is encouraged. That means minimizing the number of possible bumps/resonances. It seems like a single one for spin $0$ is the minimum.

There is also a sequence of broader zeroes. Calling them resonances or not would be debatable, much like calling the QCD $\sigma$ resonance  a particle was debatable in pion particle physics \cite{Pelaez:2015qba}. Here these zeros are really \textit{``more imaginary than real"} which would make their particle interpretation even more dubious. It would be fascinating to see how all these resonances arise (or not) from re-organizing the singularities of weakly coupled Virasoro-Shapiro as the coupling is increased; more on this below.

\subsection{Low Spin Dominance}

\begin{figure}[t]
 \centering
         \includegraphics[scale=0.30]{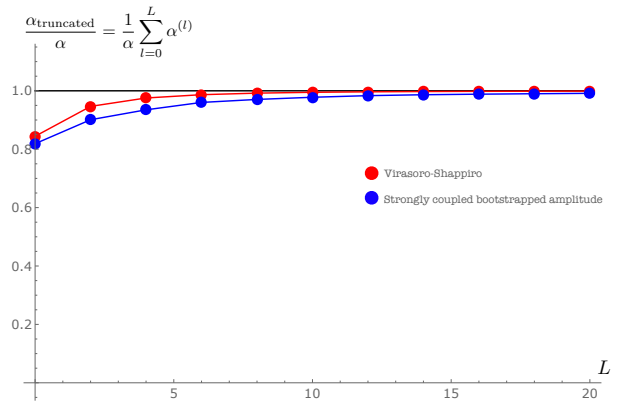}
         \vspace{-0.8cm}
 \caption{The Wilson coefficient $\alpha$ can be written as a sum rule: the integral of the imaginary part of the amplitude in the forward limit. Decomposing the latter into its various spin contributions we can assign a spin by spin contribution to the Wilson coefficient. Both the weakly coupled perturbative String amplitude as well as the strongly coupled bootstrapped amplitude (plotted here for $d=10$ at $N=30$ and $L=200$) are dominated by the lowest spins.}
 \label{fig:alphaDist}
 \end{figure}

Since $\alpha$ is also given by a sum rule (\ref{alphasumrule0}) we can partial wave decompose $A$ in the sum rule integrand to estimate how much each spin ``contributes" to $\alpha$. We find -- perhaps not surprisingly -- that lower spins contribute the most to the result as depicted in figure \ref{fig:alphaDist} for our bootstrapped strongly coupled amplitude in $d=10$. For comparison, in that same figure we plotted the contribution to $\alpha$ spin by spin in perturbative string theory where the same low spin dominance is also neatly observed despite these amplitudes being in dramatically different physical regimes. The computation of the perturbative string theory values is in appendix \ref{VSappendix}.

\section{Discussion}

\subsection{Inelasticity from Black Hole Production}

\begin{figure}[t]
 \centering
         \includegraphics[scale=0.29]{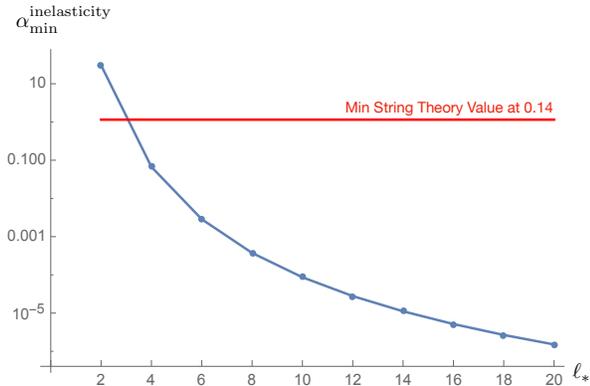}
         \vspace{-0.8cm}
 \caption{If we start around $\ell_*$ equal to $4$ or $6$ we get an order of magnitude similar to the gap between numerics and string theory. Much larger $\ell_*$ produce a negligible bound and smaller $\ell_*$'s would lead to a much too strong bound even excluding string theory. This plot is done for $d=10$; $d=9,11$ are similar (the only significant difference is that for $d=11$ both $\ell_*=2$ \text{and} $\ell_*=4$ would predict too much inelasticity; there we would need $\ell_* \ge 6$.)}
 \label{lstar}
 \end{figure}

At high energy, for each fixed $\ell$ we expect to enter the regime of black hole production eventually. Above that energy $s_*(\ell)$ we expect inelasticity to dominate and $|S_\ell(s)| \simeq 0$, see e.g. \cite{Giddings:2009gj,Haring:2022cyf}. This has implications on $\alpha$. Indeed, we know that (\ref{alphasumrule0}) 
is positive since the integrand is positive. Black hole inelasticity allows us to improve this positivity bound $\alpha>0$. Decomposing $A(s,t)$ in the sum rule into partial waves we have (here we are specializing to $d=10$ for simplicity)
\beq
 \alpha  =  \frac{16}{3\pi^4 \ell_P^{14}} \sum_\ell (\ell+1)_6(2\ell+7) \int\limits_{0}^\infty ds\, \frac{  1-\text{Re}(S_\ell(s)) }{s^8  } \,,
\label{alphasumrule01}
\eeq
If we assume that for each spin there is a minimal energy $s_*(\ell)$ after which the corresponding $S$-matrix is effectively vanishing, we get 
\beq
 \alpha \gtrsim  \frac{16}{3\pi^4 \ell_P^{14}} \sum_{\ell=\ell_*}^\infty (\ell+1)_6(2\ell+7)  \int\limits_{s_*(\ell)}^\infty  \frac{ds}{ s^8  } \,.
\label{alphasumrule02}
\eeq
To produce a lower bound on $\alpha$ we need a model for the transition value $s_*(\ell)$. The simplest estimate is to define a semi-classical impact parameter $b\equiv 2\ell/\sqrt{s}$ and set it equal to the the Schwarzschild radius   $R_S^7=\tfrac{105 \pi^3 \ell_P^8}{2} \sqrt{s}$
to get 
\beq 
 s_*(\ell) \simeq\frac{1}{\ell_P^2}\, \Big(\frac{2^{8} \ell^7}{105\pi^3}\Big)^{1/4}\,.
\eeq
Of course, for the semi-classical picture to be valid we need the spin to be large so we should sum starting at some critical $\ell_*$ in (\ref{alphasumrule02}), ignoring the contributions from the lowest spins $\ell<\ell_*$ (after all we are interested in a lower bound). As we see in figure \ref{lstar}, if $\ell_*$ is too small, this crude model would predict a lower bound way too stringent, even excluding string theory; if $\ell_*$ is too large the predicted inelasticity is negligible. Would be great if the small gap between the bootstrap and String theory values reported in table \ref{alphaResultTable} could have a simple inelasticity explanation. Properly incorporating inelasticity in the SUGRA bootstrap, taking into account the $\ell<\ell_*$, and making all this precise is an important open problem we are currently investigating together with Lance Dixon.    

\subsection{Other Wilson Coefficients and Sum Rules}
Of course, $\alpha$ is just the first out of infinitely many Wilson coefficients which parametrize the low energy expansion of any SUGRA UV completion as
\begin{align}
A(s,t)=\frac{1}{stu}&  \left[1 + \alpha \, \ell_P^6 s t u +\ell_P^{d-2}f_{1 }(s,t,u) \right. \\
 &\quad\left. + \beta \ell_P^{10} (stu)(s^2+t^2+u^2)+\cdots \right]\,,\nonumber
\label{eq:TWCs}
\end{align}
Here we see the second Wilson coefficient -- $\beta$ -- as well as the universal one loop correction $f_1(s,t,u)$ computed in appendix~\ref{unitAppendix}. These are the two leading corrections beyond~$\alpha$. 

In String theory $(\alpha,\beta)$ carve out a beautiful space as depicted in figure \ref{QGspaceCartoon} in ten dimensions. The numerical bootstrap is a convex optimization problem and $\alpha$ and $\beta$ are linear functions of the S-matrix. So, the allowed bootstrap region, must be no smaller than the convex hull of the String theory values. This is depicted by the gray region in figure \ref{QGspaceCartoon}. The lower bound $\alpha\ge0.124$ that we found is clearly away from this minimal region, however as we discussed above, we expect this bound to tighten when we input inelasticity due to black hole production. It would be fascinating to see what is the actual space predicted by the Bootstrap. Will it mimic something like the gray shape in the figure -- which could be seen as either a remarkable coincidence or as strong evidence for the uniqueness of String theory as a consistent UV completion -- or will it be a much larger region? In principle we have all the technology to settle this question: 
\begin{figure}[t]
\centering
        \includegraphics[scale=0.45]{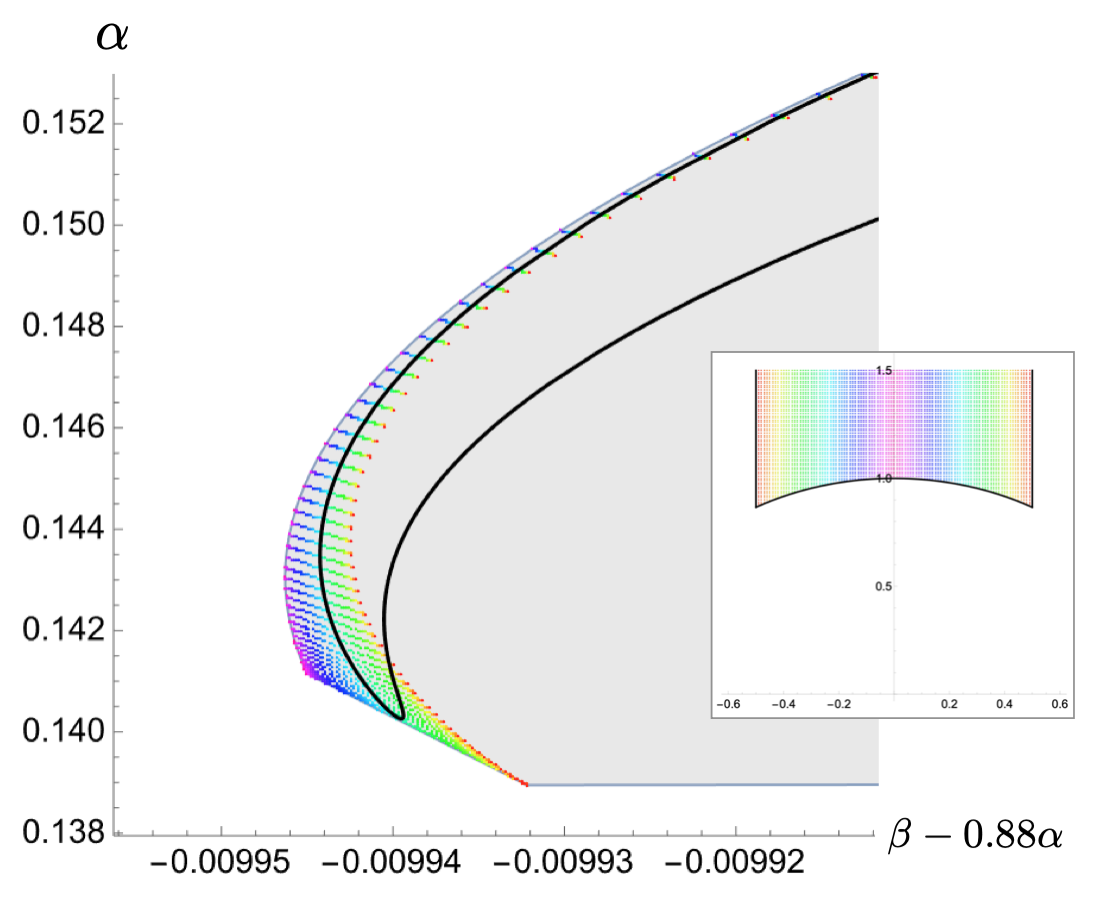}
\caption{Allowed values of $\alpha$ and $\beta$ in 10d type IIB String theory (colourful region corresponding to various moduli in the fundamental region indicated by the inset) and in type IIA String theory (solid black line). The convex hull of these allowed values is the region in gray (the lower boundary of the convex hull is horizontal because when the complex string coupling $\tau \to i\infty$ the ratio $\alpha/\beta \to 0$ in type IIB string theory.) 
Since the numerically bootstrap problem is convex the space of QG theories carved out through this problem could at most capture this region. The current numerical estimate for the minimum of $\alpha$ ($\simeq 0.124$) would be outside the range of this plot and the measured $\beta$ through arcs would be much to the left. If we trust this point as being inside the primal region then the convex hull would be much larger of course. Accounting for inelasticity it might approach the String theory region. }
\label{QGspaceCartoon}
\end{figure}
as done in the pion bootstrap \cite{Pions2}, we should add the $f_1$ threshold behavior to the ansatz so we can access the subleading coefficient $\beta$. Then we can simply hold $\alpha$ fixed and maximize/minimize $\beta$. We hope to report on progress in this direction soon.  

Here, instead of carving out this space in this more rigorous way we will discuss how we can try to measure $\beta$ for the extremal amplitudes we found which minimize $\alpha$. In short -- and with a big grain of salt since this extraction procedure is subtle -- we find that these amplitudes have a very small $\beta$. If true, these would indicate that $(\alpha_\text{min}^\text{bootstrap},0)$ is (either inside or very close being) inside the primal space.

We can read off the leading Wilson coefficients neatly by focusing on  forward limit. At $-t \ll s \ll \ell_P^{-2}$ we have that $A(s,t)-1/stu$ is approximately equal to  
\beq
\left\{ 
\begin{array}{ll}
\alpha \,\ell_P^{6} + \eta \ell_P^7 \(s^{1/2} + (-s)^{1/2}\)+ 2\beta \ell_P^{10} s^2 &,\,d=9\\
\alpha \,\ell_P^{6} + i\eta \ell_P^8 s+ 2\beta \ell_P^{10} s^2 &,\,d=10\\
\alpha \,\ell_P^{6} + \eta \ell_P^{9} \(s^{3/2} + (-s)^{3/2}\)+ 2\beta \ell_P^{10} s^2  &,\,d=11
\end{array}
\right. \la{thresholdTrue}
\eeq
where the one-loop constant $\eta$ is fixed by the corresponding $d$-dimensional $f_1$ in the true amplitude, 
\beq
\eta=\left\{
\begin{array}{lll}
\frac{\pi^3}{480} & \simeq 0.0646 & \,, d=9\\
\frac{\pi^3}{1152} & \simeq 0.0269 & \,, d=10\\
\frac{\pi^4}{8960} & \simeq 0.0109 & \,, d=11
\end{array}
\right.
\label{etaValsAlld}
\eeq
After all unitarity is saturated at threshold up to two loops.

\begin{figure}[t]
    \centering
    \includegraphics[width=\linewidth]{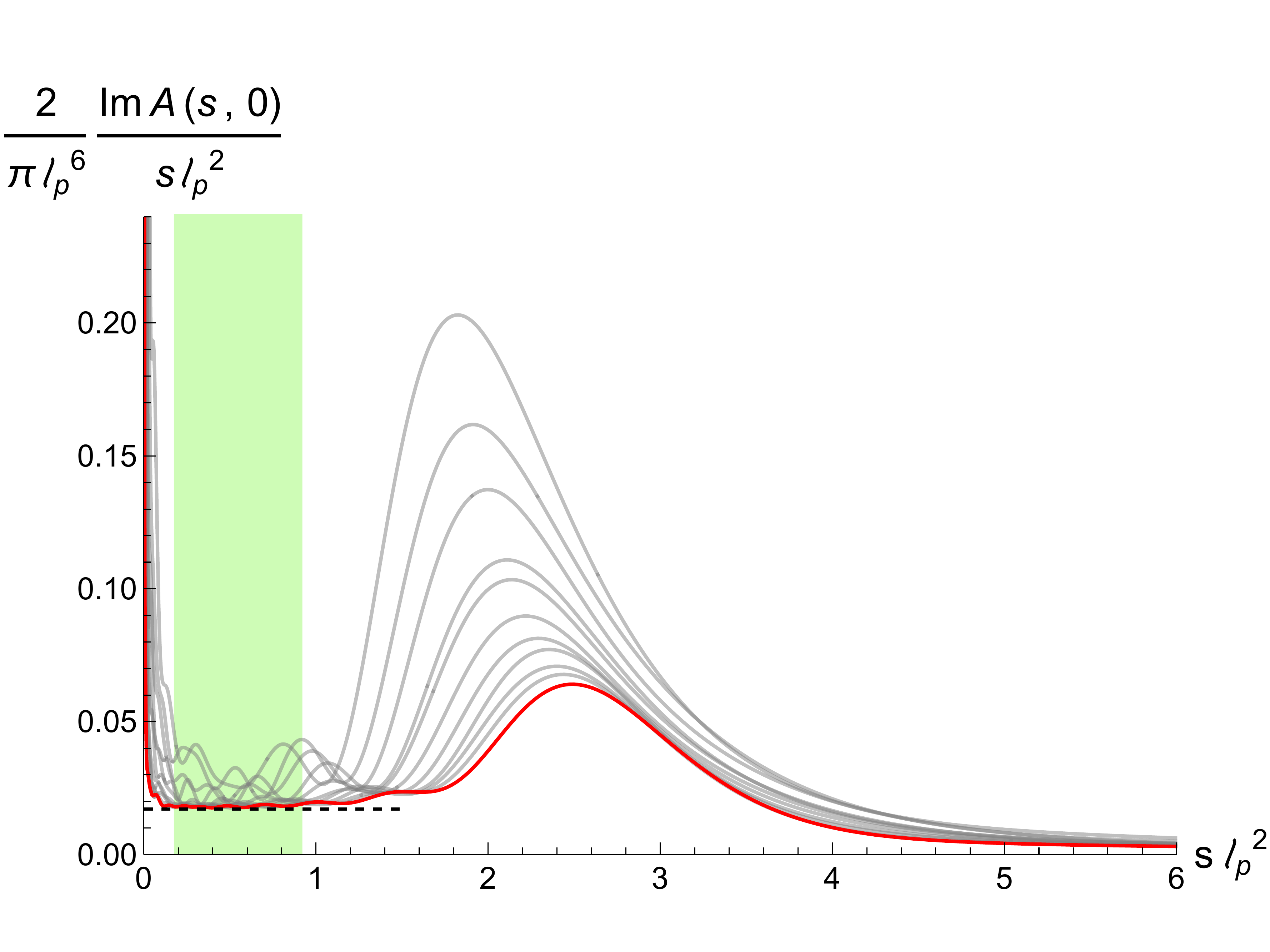}
    \caption{Integrand in the sum rule for $\alpha$. The different lines correspond to $N=20$ to $30$ and with the red line being $N=30$. The horizontal dashed line is the one-loop term which is well approximated in the green region by the ansatz. Therefore, eqns.\,(\ref{10dSumRules}) hold in this region.}
    \label{fig:integrandAlpha}
\end{figure}

Note that our $\rho$ series ansatz will, strictly speaking, \textit{not} have this behavior. In $d=10$, for instance, we see from (\ref{thresholdTrue}) that we should have a logarithmic behavior while the $\rho$ ansatz contains only square roots! This is not a huge deal, with an infinite $\rho$ series we can reproduce any analytic function. In figure \ref{fig:integrandAlpha} we see how the constant imaginary plateau (from ${\rm Im}\, \log(-|x|) = i \pi $) is indeed better and better approached by the $\rho$ series as $N$ increases. As this plot illustrates, this approach is not point-wise. The behavior very close to the origin is always a bit polluted and as we go to larger energies, the EFT description breaks down due to the graviball resonance. There is a nice intermediate region depicted in green -- which we colloquially call the EFT region -- where this constant behavior is well captured and the height matches the perturbative expectation. What we can then do is to consider a point at a small $s_0$ in the EFT region so that we can extract all constants $(\alpha,\beta,\eta)$ from three simple sum rules
\begin{subequations}
\begin{align}
    \alpha - \frac{2\eta}{\pi} \ell_P^2 s_0 &\approx \frac{2}{\pi\ell_P^6} \int\limits_{s_0}^\infty ds\, \frac{ {\rm Im}\, A }{s}, \\
    -\eta + \frac{2\alpha}{\pi\ell_P^2s_0} - \frac{4\beta}\pi \ell_P^2s_0&\approx 
    \frac{2}{\pi\ell_P^8} 
    \int\limits_{s_0}^\infty ds\, \frac{ {\rm Re}\, A }{s^2}, \\
    \beta +  \frac{\eta}{\pi\ell_P^2s_0} &\approx 
    \frac{1}{\pi\ell_P^{10}} 
    \int\limits_{s_0}^\infty ds\, \frac{ {\rm Im}\, A }{s^3},
\end{align}
\label{10dSumRules} 
\end{subequations}
where in the right hand side we plug the optimal amplitudes obtained from our numerics evaluated at $t=0$ and $s$ infinitesimally above the real axis. What we did -- see appendix \ref{ap:Arcs} -- was a bit better than this. For several $N$'s, we considered several $s_0$ in the EFT region and fitted the resulting sum rules to extract the best $(\alpha,\beta,\eta)$ which we then extrapolated to $N=\infty$. There are two important cross-checks of this procedure: 
\begin{itemize}
    \item $\eta$ better agree with the analytic prediction from the one loop correction computed in appendix \ref{unit1Loop}.  
    \item $\alpha$ should match with the $\alpha$ we got from the minimization.
    \end{itemize}
While this nicely works not only in $d=10$ but also in the other $d=9,11$ -- see panels in appendix \ref{ap:Arcs} -- the $\beta$ fits are quite erratic and therefore we are unable to get reliable fits. However, they appear to converge to values smaller than the String Theory predictions as anticipated above. 

Again, the rigorous thing to do is to impose the right threshold behavior in the ansatz and extremize both $\alpha$ and $\beta$.\footnote{In $d=9$ the behavior of the ansatz is already a bit closer to the real behavior and indeed the arc fits seem to agree much better with the extremization targets.} We will do it soon.

{\begin{center}{\textbf{ACKNOWLEDGMENTS}} \end{center}}
We are grateful to Francesco Aprile, Nathan Berkovits, Lance Dixon, Michael Green and Simon Caron-Huot for numerous enlightening discussions. We also thank Simon for useful comments on the draft, and for pointing out the importance of a double scaling limit of large energy and spin. 
Research at the Perimeter Institute is supported in part by the Government of Canada through NSERC and by the Province of Ontario through MRI. This work was additionally supported by   grants from the Simons Foundation (PV: \#488661, JP: \#488649) and from FAPESP  (2016/01343-7 and 2017/03303-1). 
The work of 
JP is supported   by the Swiss National Science Foundation through the project
200020\_197160 and through the National Centre of Competence in Research SwissMAP. 
AG is supported by the European Union - NextGenerationEU, under the programme Seal of Excellence@UNIPD, project acronym CluEs.
This work was also 
supported by The Israel Science Foundation (grant number 2289/18), and 
in part by the National Science Foundation under Grant No. NSF PHY-1748958.

\appendix
\section{Decompactification Limits}
\label{reductions}

In string theory, the values of $\alpha$ in $d+1$ spacetime dimensions can be obtained from $\alpha$ in $d$ dimensions by taking the decompactification limit $R\to \infty$, where $R$ is the radius of the compact extra dimension. This provides a consistency check of the formulas quoted in section \ref{alphaST}.

Firstly, notice that we can relate the gravitational constant across dimensions by comparing the coefficient of the Ricci scalar in the spacetime action. This gives $16\pi G_{d}  = 16\pi G_{d+1}/(2\pi R)$, which leads to 
\beq
(\ell_d)^{d-2} R = (\ell_{d+1})^{d-1}\,,
\label{ellPdimensions}
\eeq
where $\ell_d$ denotes the Planck length in $d$ spacetime dimensions.
Then, from the scattering amplitude \eqref{defAlpha}, we expect
\beq
\lim_{R\to \infty} \alpha_d \ell_d^6 =\alpha_{d+1} \ell_{d+1}^6\,.
\eeq

\noindent\rule[0.5ex]{\linewidth}{1pt}

\noindent $\mathbf{IIA\ d=10}  \longrightarrow \mathbf{d=11}.$
Using \eqref{ellPdimensions} and the expression of the type IIA string coupling $g_s = (R/\ell_{11})^{\frac{3}{2}}$ in terms of the radius of the M-theory circle and the 11-dimensional Planck length, it is straightforward to check that 
\beq
\lim_{R\to \infty} \alpha^\text{IIA} \ell_{10}^6 =\frac{\pi^2}{96}  \ell_{11}^6\,,
\eeq
 in agreement with \eqref{alphaMtheory}.

\noindent\rule[0.5ex]{\linewidth}{1pt}

\noindent $\mathbf{IIA\ d=9}  \longrightarrow \mathbf{d=10}.$
In order to check this decompactification limit, one needs the following formulas
\beq
\nu=\left(\frac{R}{\ell_{10} }\right)^\frac{3}{2} g_s^{\frac{7}{8}} \,,
\qquad
{\rm Im}\,\tau = \frac{R}{\ell_{10} } g_s^{-\frac{3}{4}}\,,
\eeq
that relate the 9 dimensional parameters to the 10 dimensional ones \cite{Pioline:2015yea}.\footnote{These formulas follow from the ones in \cite{Pioline:2015yea} using the standard relations $\ell_{10}^8 = g_s^2 \ell_s^8$ and $\ell_9^7 = g_9^2 \ell_s^7$ that involve the string length $\ell_s$ and the 9-dimensional coupling $g_9$.}
Then, using $E_\frac{3}{2}(\tau,\bar{\tau})\sim 2\zeta_3 
({\rm Im}\,\tau)^{\frac{3}{2}}$ for large ${\rm Im}\,\tau$, we can easily check that
\beq
\lim_{R\to \infty} \frac{\ell_9^6}{2^6}\left[ \nu^{-\frac{3}{7}} E_\frac{3}{2}(\tau,\bar{\tau}) +  \frac{2\pi^2}{3}\nu^\frac{4}{7}\right]=  \ell_{10}^6 \alpha^\text{IIA} \,.
  \eeq

\noindent\rule[0.5ex]{\linewidth}{1pt}

\noindent $\mathbf{IIB\ d=9}  \longrightarrow \mathbf{d=10}.$ 
In this case, we keep fixed the complexified string coupling $\tau$ and the 10-dimensional Planck length $\ell_{10}$, and use $\nu = (\ell_{10}/R)^2$ and   \eqref{ellPdimensions} to compute the limit:
\beq
\lim_{R\to \infty} \frac{\ell_9^6}{2^6}\left[ \nu^{-\frac{3}{7}} E_\frac{3}{2}(\tau,\bar{\tau}) +  \frac{2\pi^2}{3}\nu^\frac{4}{7}\right]= \frac{\ell_{10}^6}{2^6} E_\frac{3}{2}(\tau,\bar{\tau})\,.
  \eeq

\section{Numerical Setup} \label{ap:numerics}

The numerical optimization follows \cite{ourST} and uses most of the same technology. To start with, we have the ansatz (\ref{ansatz0}). In order to have the correct large energy and large spin behaviour, we impose the linear constraints described in Appendices G and F of \cite{ourST}. We impose the unitarity conditions eq.\,(\ref{Sunit}) as SDP constraints for $\ell=0,2,4\ldots L$ and $s\in [0,\infty)$ on a Chebyshev grid of size 326.

Additionally, for each grid point in $s$, we impose the novel \textit{Positivity in the Sky} conditions with SDP matrices of size $d\times d$ in d-dimensions. As we increase the energy, the impact parameter decreases and we need to shrink the opening angle $\theta_0$ of the cone (Figure \ref{fig:positivity}) in order to see non trivial scattering. Therefore, we scale $\theta_0$ such that $t_0=\frac{-s}2(1-\cos\theta_0)=-0.005$ is held fixed.

In addition to the positivity conditions considered in the main text, we can also do something intermediate between full \textit{Unitarity in the Sky} and linear \textit{Positivity in the Sky}. Following the split in (\ref{ansatz0}) into the SUGRA tree-level part \texttt{tree} and the rest -- which we will refer to as \texttt{loops} -- we can split 
\beq
2\,\text{Im} \mathbb{t} - \mathbb{t}_\texttt{tree}^\dagger\mathbb{t}_\texttt{tree}-\mathbb{t}_\texttt{tree}^\dagger\mathbb{t}_\texttt{loops}-\mathbb{t}_\texttt{loops}^\dagger\mathbb{t}_\texttt{tree} \succeq \mathbb{t}_\texttt{loops}^\dagger\mathbb{t}_\texttt{loops} \succeq 0  \nn
\eeq
and impose that the left hand side is positive semi-definite, see figure \ref{tikzpicture}. This still requires computing the integrals over the intermediate phase space when multiplying \texttt{tree} and \texttt{loops} but those are done once and for all and the left hand side remains linear in the \texttt{loops} variables so the $M^2$ cost is gone. This extra constraint is \textit{not} stronger (or weaker) than positivity since the mixed terms in the left hand-side do not have a definite sign. Instead, it is an extra condition we can easily impose. We call it  {\textit{Tree-level Subtracted Positivity in the Sky}}. However, we checked that these extra constraints do nothing significant in the examples studied in this paper; their tiny effects can not even be seen in our plots.

\begin{figure}
    \centering
    \begin{tikzpicture}
        \begin{scope}[shift={(-3.2,0)},scale=0.8]
        \filldraw[cyan!15, opacity=0.9] (0,0) circle (0.5);
        \node[text width=0.2cm] at (0,0) {$\mathbb{t}$};
        \draw[line width=0.5mm] (0.353553,0.353553) -- (3/4,3/4);
        \draw[line width=0.5mm] (-0.353553,0.353553) -- (-3/4,3/4);
        \draw[line width=0.5mm] (0.353553,-0.353553) -- (3/4,-3/4);
        \draw[line width=0.5mm] (-0.353553,-0.353553) -- (-3/4,-3/4);
        \end{scope}
        \begin{scope}[shift={(-0,0)},scale=0.8]
        \filldraw[cyan!15, opacity=0.9] (0,0) circle (0.5);
        \node[text width=0.2cm] at (-0.15,0) {$\mathbb{t}_{\texttt{tree}}$};
        \draw[line width=0.5mm] (0.353553,0.353553) -- (3/4,3/4);
        \draw[line width=0.5mm] (-0.353553,0.353553) -- (-3/4,3/4);
        \draw[line width=0.5mm] (0.353553,-0.353553) -- (3/4,-3/4);
        \draw[line width=0.5mm] (-0.353553,-0.353553) -- (-3/4,-3/4);
        \end{scope}
        \begin{scope}[shift={(1.3,0)},scale=0.8]
        \filldraw[cyan!15, opacity=0.9] (0,0) circle (0.5);
        \node[text width=0.2cm] at (-0.15,0) {$\mathbb{t}^\dagger_{\texttt{loops}}$};
        \draw[line width=0.5mm] (0.353553,0.353553) -- (3/4,3/4);
        \draw[line width=0.5mm] (-0.353553,0.353553) -- (-3/4,3/4);
        \draw[line width=0.5mm] (0.353553,-0.353553) -- (3/4,-3/4);
        \draw[line width=0.5mm] (-0.353553,-0.353553) -- (-3/4,-3/4);
        \end{scope}
        \begin{scope}[shift={(-1,-1.5)}]
        \begin{scope}[shift={(-0,0)},scale=0.8]
        \filldraw[cyan!15, opacity=0.9] (0,0) circle (0.5);
        \node[text width=0.2cm] at (-0.15,0) {$\mathbb{t}_{\texttt{tree}}$};
        \draw[line width=0.5mm] (0.353553,0.353553) -- (3/4,3/4);
        \draw[line width=0.5mm] (-0.353553,0.353553) -- (-3/4,3/4);
        \draw[line width=0.5mm] (0.353553,-0.353553) -- (3/4,-3/4);
        \draw[line width=0.5mm] (-0.353553,-0.353553) -- (-3/4,-3/4);
        \end{scope}
        \begin{scope}[shift={(1.3,0)},scale=0.8]
        \filldraw[cyan!15, opacity=0.9] (0,0) circle (0.5);
        \node[text width=0.2cm] at (-0.15,0) {$\mathbb{t}^\dagger_{\texttt{tree}}$};
        \draw[line width=0.5mm] (0.353553,0.353553) -- (3/4,3/4);
        \draw[line width=0.5mm] (-0.353553,0.353553) -- (-3/4,3/4);
        \draw[line width=0.5mm] (0.353553,-0.353553) -- (3/4,-3/4);
        \draw[line width=0.5mm] (-0.353553,-0.353553) -- (-3/4,-3/4);
        \end{scope}
        \end{scope}
        \node[text width=1.5cm] at (-4.4, 0) {\large{2 Im} $\Biggl[$};
        \node[text width=1.5cm] at (2.8, 0) {$\large{\Biggr]}$};
        \node[text width=1.9cm] at (-1.5, 0) {\large $\Biggr]- \text{2 Re}\Biggl[$};
        \node[text width=1.5cm] at (-2.3, -1.5) {\large $-\text{ Re}\Biggl[$};
        \node[text width=1.5cm] at (1.8, -1.5) {\large$\Biggr] \succeq 0$};
    \end{tikzpicture}
    \caption{Tree-level Subtracted Positivity in the Sky 
    }
    \label{tikzpicture}
\end{figure}

\section{Unitarity in the Regge Limit} \la{runawayappendix}

At high energies and large impact parameters, the gravitational S-matrix is governed by semiclassical physics. In this regime, the tree level graviton exchange must eikonalize \cite{tHooft:1987vrq, Amati:1987wq, DiVecchia:2019kta,Haring:2022cyf}, 
\beq
 S(s,b) \approx e^{i\delta^{(0)}(s,b)},
\eeq
where $b=2\ell/\sqrt{s}$ is the impact parameter and $\delta^{(0)}$ is the tree level phase shift (\ref{EikonalPhase}). In a physical theory, this behaviour results from an infinite sum of ladder and crossed ladder diagrams \cite{DiVecchia:2019kta}, which are the leading contributions in this limit. In addition to fixing high energy unitarity, these terms lead to infinitely many resonances in the crossed channel organized into a Regge trajectory as in the Virasoro-Shapiro amplitude. However, at finite $N$, our ansatz can only have a finite number of zeros. We should therefore expect trouble with unitarity in this semiclassical regime. Indeed, as we increase $N$, we see more and more zeros entering the physical sheet as described in Appendix \ref{ap:resonances}.

More concretely, we can address this problem by looking at unitarity in the double scaling limit of large $s$ and large $\ell$ with $b^2=4\ell^2/s$ held fixed. Let us denote the partial wave amplitudes by $h_\ell(s) = -i(S_\ell(s)-1)$. The contribution of the graviton pole to the partial waves can be computed straightforwardly and is given by,
\begin{align}
    h^\text{pole}_\ell(s)&=\frac{\Gamma\left(\frac{d}{2}-2\right)}{2^{d-1}\pi^{\frac{d-2}2}}\frac{s^{\frac{d-2}{2}}}{(\ell +1)_{d-4}}\ .
\end{align}
Clearly, when $s\sim\ell^2$, this term grows like $\ell^2$. The $\rho$ ansatz must unitarize the amplitude by cancelling this divergence. We can obtain the partial wave decomposition of the ansatz in the double scaling limit using the Froissart-Gribov equation 
\begin{equation}\label{pwDef}
    h_\ell(s)= \frac{2\mathcal{N} s^{\frac{d-4}2}}\pi \int_1^\infty \dd z\ (z^2-1)^{\frac{d-4}2} Q_\ell^d(z) \mathrm{Disc}_z T(s,z)\ ,
\end{equation}
where $\mathcal{N}$ is the same normalization constant as in (\ref{SlDef}) and $Q_\ell^d(z)$ is the Gegenbauer-Q function. It is given by the following expression \cite{Correia:2020xtr},
\begin{equation}\la{qFunction}
   Q_\ell^d(z)= \tfrac{\mathcal{C}_\ell^d\ \lambda (z)^{-\ell }}{\left(\lambda (z)^2-1\right)^{\tfrac{d-3}{2}}} \, _2F_1\left(\tfrac{5-d}{2},\tfrac{d-3}{2};\tfrac{d+2\ell-1}{2} ;\tfrac{1}{1-\lambda
   (z)^2}\right),
\end{equation}
where $\lambda(z)=z+\sqrt{z^2-1}$ and $\mathcal{C}_\ell^d$ is a constant\footnote{$\mathcal{C}_\ell^d=\frac{\sqrt{\pi} \Gamma(\ell+1)\Gamma(\frac{d-2}{2})}{2^{\ell+1}\Gamma(\ell+\frac{d-1}{2})}$.}. In order to study the large spin behaviour, we need to zoom in close to the forward limit. Let us change variables to $z=1+\frac{\theta^2}{2\ell^2}$. Now, consider the following integral representation for the Hypergeometric function,
\begin{equation}
    _2F_1(a,b;c;x)=\frac{\Gamma (c) \int_0^1 \dd t\ t^{b-1} (1-t x)^{-a} (1-t)^{-b+c-1}}{\Gamma (b) \Gamma (c-b)}
\end{equation}
which is valid for $c>b>0$. Note that at large spin $\lambda(z)\approx 1$, so the argument of the Hypergeometric is very large. Therefore, the integral is dominated by the region near $t=1$. Let us again change variables to $t=1-r/\ell$ and expand the integrand in $1/\ell$. Integrating the resulting series, we get the following approximation for $\ell\gg1$,
\begin{align}
  \mathcal{N}(z^2-1)^{\frac{d-4}2} Q_\ell^d(z) &\sim \frac{(2\theta)^{\frac{d-4}{2}} \Gamma \left(\frac{d-2}{2}\right) K_{\frac{d-4}{2}}(\theta )}{\ell^{d-4} }+\ \ldots
\end{align}
where once again, $z=1+\frac{\theta^2}{2\ell^2}$. One can systematically go to higher orders in $1/\ell$. 

Plugging in a monomial of the form $\rho^i(s)\rho^j(t)\rho^k(u)$ and the above expansion for the Q-function into (\ref{pwDef}) with $s=4\ell^2/b^2$, we obtain a complicated expression which schematically looks like
\begin{equation}\la{uglyRhoExp}
    \begin{split}
    h^{ijk}_\ell\(s\)\sim &\ \mathbb{K}_0(b)\left(\ell^6 \text{poly}(1/b) + \ell^5\text{poly}(1/b) + \ldots\right) \\
    &+ \mathbb{K}_1(b)\left(\ell^6\text{poly}(1/b) + \ell^5\text{poly}(1/b) + \ldots\right)\\
    &+ \(\ell^6\text{poly}(1/b) + \ell^5\text{poly}(1/b) +\ \ldots\), 
    \end{split}
\end{equation}
where poly$(1/b)$ is a placeholder for polynomials in $1/b$ whose degree depends on $i,j,k$ and $\mathbb{K}_\nu(x)$ are Struve functions of the second kind. This expression grows like $\ell^6$, so it might be possible to unitarize the pole term which grows like $\ell^2$.

\begin{figure}[t]
    \centering
    \includegraphics[scale=0.46]{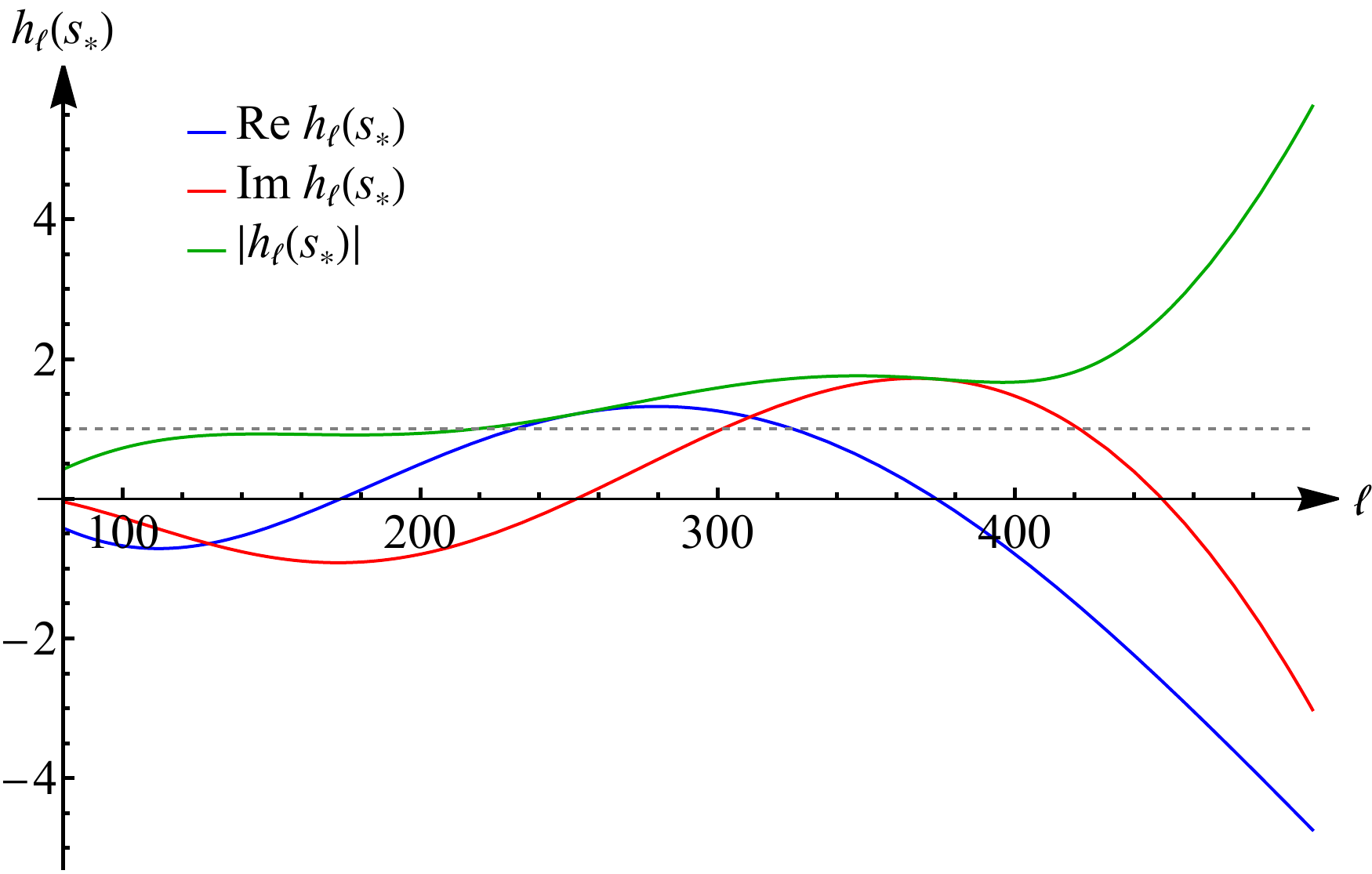}
    \caption{Here we plot $h_\ell(s_*)$ for a fixed $s_*\approx 7\times10^{10}$ up to corrections of order $1/\ell^3$ for $N=24$ and $L=200$ in 10d. The dashed grey line is unity. We have imposed full unitarity up to $\ell=200$ and interestingly, there are unitarity violations for $\ell> 200$.}
    \label{fig:unitViolationRegge}
\end{figure}

We need to cancel the coefficients of positive powers of $\ell$ in $\(h_\ell^\text{pole}(s) + \sum_{ijk}\alpha_{ijk}h^{ijk}_\ell(s)\)$. We need unitarity to hold at all energies, and therefore at all $b$. Also, the terms in the three lines of (\ref{uglyRhoExp}) cannot mix. So, we need to cancel separately all the structures that come with different powers of $b$ in poly$(1/b)$. This gives us a set of linear constraints analogous to the high energy fixed angle conditions in \cite{ourST}. It turns out however that at finite $N$ there are no non-trivial solutions (i.e. with non-zero $G_N$). Therefore we must encounter infeasibility as we go to very large $L$ at fixed $N$.

In practice, the bootstrap works hard to delay this divergence to spins larger than $L$. To see this better, consider the extremal amplitude for $N=24$ and $L=200$. We computed the partial waves $h_\ell(s)$ in the manner described above upto order $1/\ell^2$ and evaluated it at a large value of $s_*\approx 7\times 10^{10}$. In figure \ref{fig:unitViolationRegge}, we see that there are unitarity violations for $\ell>L=200$. This means that the numerical optimization procedure pushes the violations to just beyond the point where we are probing it. Other values of $N$ and $L$ display the same behavior. 

\begin{figure}[t]
    \centering
    \includegraphics[scale=0.46]{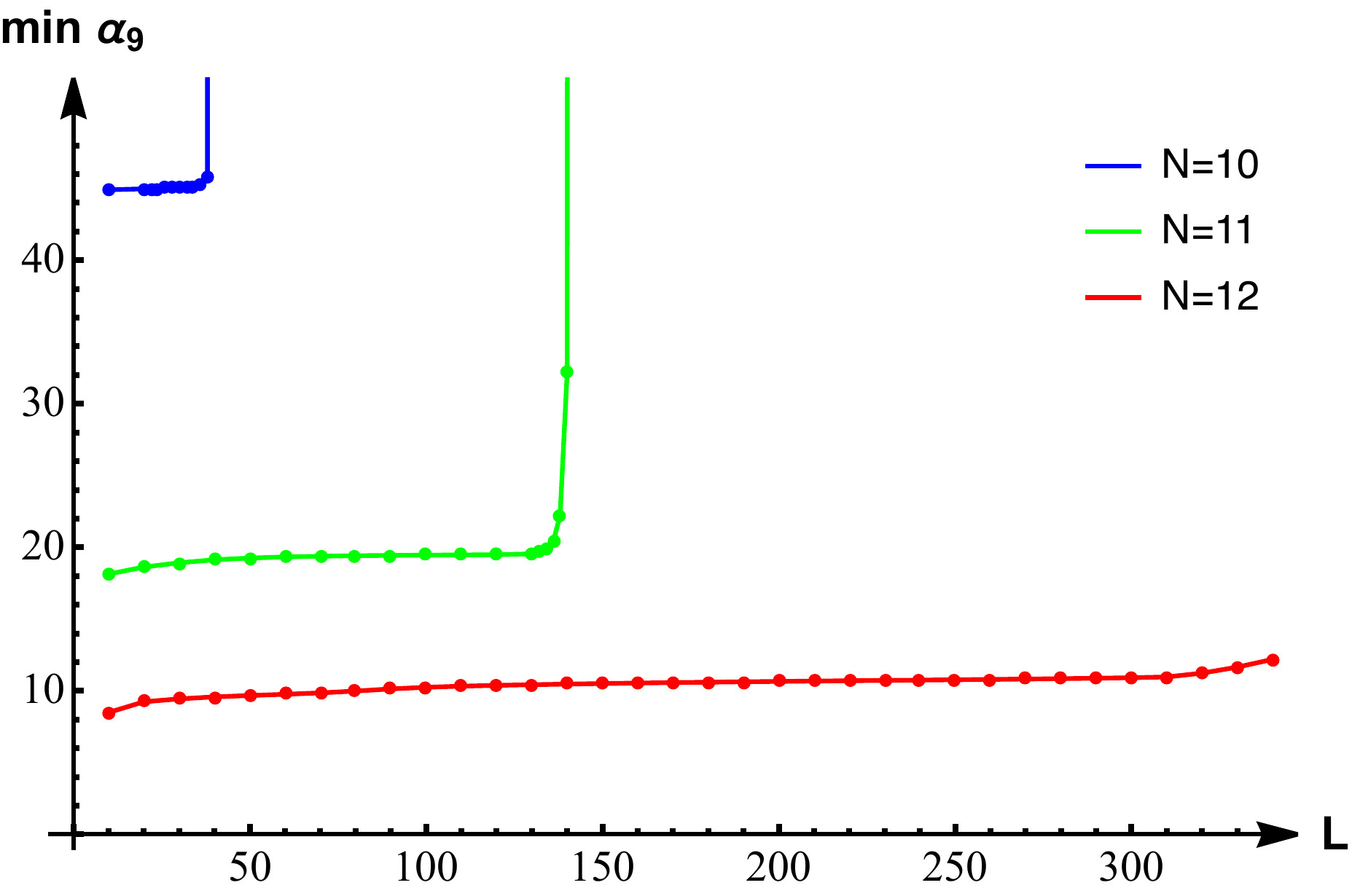}
    \caption{For large enough energy and spin we will run into trouble with unitary at any finite $N$. The smallest value of $N$ for which the high energy linear constraints from \cite{ourST} are feasible is $N=10$. We see here that for $N=10$ and $11$, we encounter infeasibility as we impose more unitarity conditions at $L=50$ and $L=150$ respectively. For $N=12$ we begin to lose the plateau near $L=340$. This is as far as we went in spin, but it seems likely that $N=12$ will become infeasible for $L\approx 360$. Since the runaway will be more and more delayed with larger $N$, to be more rigorous and avoid this trouble, we could do a fit in $L$ and $N$ at the same time rather than extrapolating in $L$ first and $N$ last. Such mixed fits were explored in \cite{Haring:2022sdp}.}
    \label{plateauRunaway}
\end{figure}

This analysis tells us that the apparent plateaus in figure \ref{ResultsAllD} must be transient and eventually all those curves need to shoot up. This infeasibility will be more pronounced for small values of $N$ because the ansatz has less freedom to tame these effects, see figure \ref{plateauRunaway}. Of course, as $N \to \infty$ these plateaus will be longer and longer such that fitting them and then extrapolating in $N$ should yield the optimal minimal value of $\alpha$. For the range of $N$ considered in figure \ref{ResultsAllD}, we expect the infeasibility to show up at very high spins beyond the reach of our numerics. The naive extrapolations to $L \to \infty$ in section \ref{results} should always be understood as estimates for the heights of these long transient plateaus.

Is it possible to invent a better (crossing symmetric and analytic) ansatz that can satisfy unitarity at all energies and for all spins, with a finite number of parameters?
This is a very interesting question that we do not know the answer.
 
\section{One Loop Unitarization}
\la{unitAppendix}
Here we determine the one-loop contribution to \eqref{defAlpha} using elastic unitarity. This appendix complements appendix B of \cite{ourST} where this was carried out in detail in $d=10$ and where the general $d$ setup was described. As explained there, the main relation in this computation reads 

\beq
{\rm Im}\, f_1 = \frac{8s^{\frac{d}{2}-3} t u}{(2\pi)^{d-3}}    \sum_{\ell=0 \atop even}^\infty \left( \delta_\ell^{(0)}\right)^2
P_\ell^{(d)}\left(\frac{u-t}{u+t}\right)   \label{unit1Loop}
\eeq
which can be summed in closed form in a given spacetime dimension. Here $\delta_l^{(0)}$ is the tree level phase shift 

\begin{align}\la{EikonalPhase}
\delta_\ell^{(0)} =  \frac{\pi^{\frac{d-3}{2}} \Gamma(d-4)   }{2^{d-1}\Gamma\left( \frac{d-3}{2} \right)  (\ell+1)_{d-4}}  \,.
\end{align}
Once we find the imaginary part from (\ref{unit1Loop}) we find the full one loop amplitude $f_1$ in 
 \beq
 T=
\frac{(2\pi\ell_P)^{d-2}}{4\pi}\frac{s^4}{stu}  \left(1 + \alpha \, \ell_P^6 s t u +
 \ell_P^{d-2}f_{1 }(s,t,u) +\cdots \right)
 \label{eq:Tscalars}
 \eeq
by finding an appropriate crossing symmetric expression with that imaginary part. 

\noindent\rule[0.5ex]{\linewidth}{1pt}
\noindent $\mathbf{d=10}.$
\begin{align}
\label{Imf1}
{\rm Im}\, f_1 =
\frac{\pi ^3 s^3 }{7680 t^2 u^2}
   &\left[-s t u
   \left(s^2+7 (t-u)^2\right) \right.\\
   &\left.+8 t^5
   \log \left(-t/s\right)+8
   u^5 \log
   \left(-u/s\right)\right]\,.\nonumber
   \end{align}
From this one can use analyticity and crossing to reconstruct the function $f_1$ from its imaginary part,
\begin{align}
&f_1= \frac{\pi ^2 }{960}\left[
   -\frac{s^3 \left(t^5+u^5\right)  \log^2(-s)}{2 t^2u^2}\right. \label{finalf1}
 \\&
   -\frac{t^3 \left(s^5+u^5\right) \log ^2(-t)}{2 s^2 u^2}   -\frac{u^3\left(s^5+t^5\right) \log ^2(-u)}{2s^2 t^2}
     \nonumber \\& \nonumber
   +\frac{s^3 t^3 \log (-s)\log (-t)}{u^2}
   +\frac{s^3 u^3 \log(-s) \log (-u)}{t^2} \\&\nonumber
   +\frac{t^3 u^3\log (-t) \log (-u)}{s^2}
   -\frac{s^4   (s^2+7(t-u)^2 ) \log (-s)}{8 t u}
   \\&\nonumber
   -\frac{t^4 (t^2+7(s-u)^2 ) \log (-t) }{8 s u}
   -\frac{u^4  (u^2+7  (s-t)^2 ) \log ({-}u)}{8 s  t}\\&\nonumber
   \left.
   +\frac{1}{8}
   \left(s^2+t^2+u^2\right)^2
   \Big(1+\frac{\pi ^2 \left(s^6+t^6+u^6-13 s^2 t^2 u^2\right)}{2 s^2
   t^2 u^2}   \Big)
   \right]
\end{align}
One can easily check that this crossing symmetric function has the s-channel imaginary part given by \eqref{Imf1}. The rational part is  fixed by dimensional analysis and  the requirement that $f_1$ vanishes at $t=0$, so that the residue of the $T$ amplitude at $t=0$ is not affected by the one loop contribution. 

\noindent\rule[0.5ex]{\linewidth}{1pt}
\noindent $\mathbf{d=11}.$
\begin{align}
{\rm Im}\, f_1=&\frac{\pi^4  s^{9/2} \left(23 t^2 u^2-20 t^3 u-20 t u^3+15 t^4+15 u^4\right)}{46080 t^2 u^2}\nonumber\\
&-\frac{\pi ^5 s^{7/2} \sqrt{t u}
   \left(t^6+u^6\right)}{12288 t^3 u^3}\\
   &+\frac{i \pi^4  s^{7/2} \sqrt{t u} \left(u^6-t^6\right) \log \left(\frac{\sqrt{-u}+i
   \sqrt{-t}}{\sqrt{-t}+i \sqrt{-u}}\right)}{6144 t^3 u^3}\nonumber
\end{align}
This matches the result in \cite{Alday:2020tgi} (up to an overall normalization),
which also gives the full result 
\begin{align}
f_1=\frac{\pi^4  s t u }{46080  }\left(
B(s,t)+B(s,u)+B(u,t)
 \right)
\end{align}
where
\begin{align}
&B(s,t)=  
   -\frac{15 (s t)^{\frac{5}{2}} \log \left(\frac{\left(\sqrt{-s-t}+\sqrt{-s}\right)
   \left(\sqrt{-s-t}+\sqrt{-t}\right)}{\sqrt{s t}}  \right)}{(-s-t)^{7/2}}\nonumber  \\
  &- \frac{(-s)^{\frac{5}{2}} \left(3 s^2+11 s t+23 t^2\right)+(-t)^{\frac{5}{2}} \left(23 s^2+11 s t+3 t^2\right)}{(s+t)^3}
   \end{align}
\noindent\rule[0.5ex]{\linewidth}{1pt}
\noindent $\mathbf{d=9}.$
\begin{align}
{\rm Im}\, f_1=&-\frac{\pi ^3 s^{7/2} \left(3 t^2-4 t u+3 u^2\right)}{768 t u}
\nonumber\\
&+\frac{\pi ^4 s^{5/2} \sqrt{t u} \left(t^4+u^4\right)}{1024 t^2
   u^2}\\
   &+\frac{i \pi ^3 s^{5/2} \sqrt{t u} \left(t^4-u^4\right) \log \left(\frac{\sqrt{-u}+i \sqrt{-t}}{\sqrt{-t}+i
   \sqrt{-u}}\right)}{512 t^2 u^2}\,.\nonumber
\end{align}
This leads to 
 \begin{align}
f_1= 
B(s,t)+B(s,u)+B(u,t)
\end{align}
where
\begin{align}
&B(s,t)=  
  \frac{\pi ^3 (s t)^{5/2} \log
   \left(\frac{\left(\sqrt{-s-t}+\sqrt{-s}\right) \left(\sqrt{-s-t}+\sqrt{-t}\right)}{\sqrt{s t}}\right)}{256 (-s-t)^{3/2}}
   \nonumber  \\
  &-\frac{\pi ^3 s^3 \sqrt{-t} (2 s+5 t)}{768 (s+t)}-\frac{\pi ^3 \sqrt{-s} t^3 (5 s+2 t)}{768 (s+t)} \,.
   \end{align}

\begin{figure}[t]
\centering
        \includegraphics[scale=0.33]{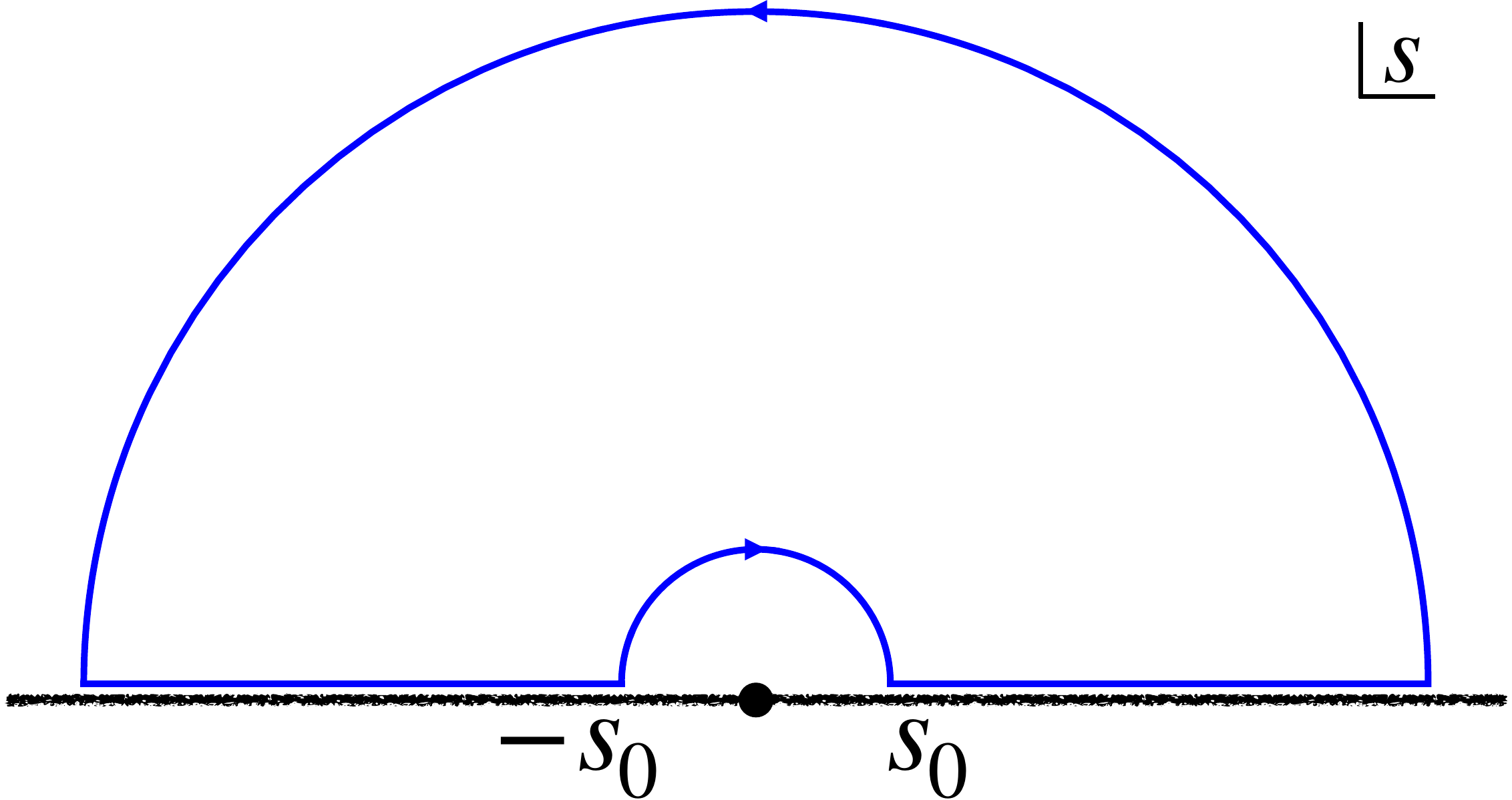}
\caption{ Integration contour leading to equation \eqref{idCauchy}. The large arc does not contribute because $f(s)\to 0$  when $|s|\to \infty$.}
\label{fig:arcs}
\end{figure}

\begin{figure*}[t]
\centering
        \includegraphics[scale=0.58]{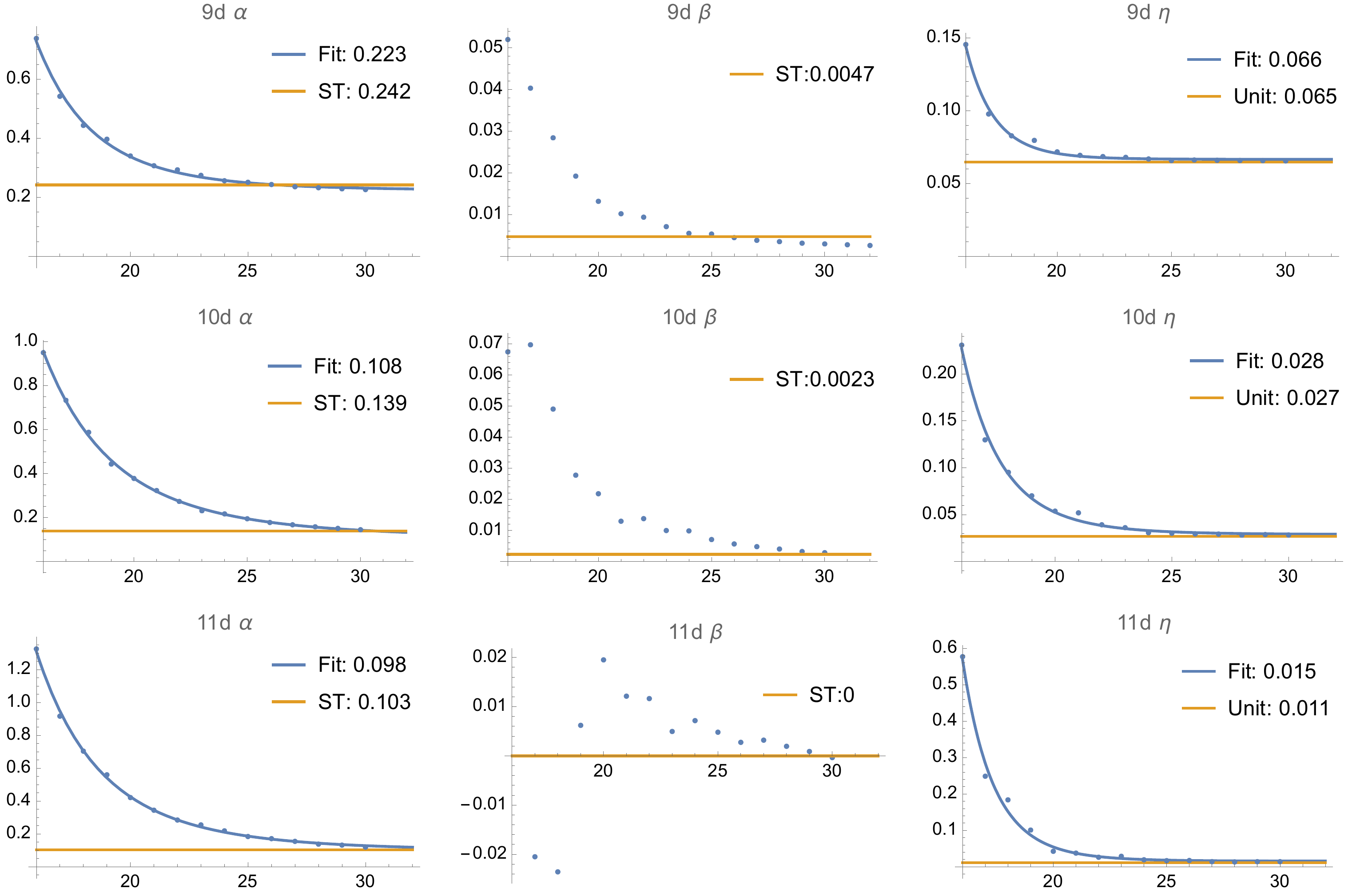}
\caption{In all dimensions, $\alpha$ extracted from the sum rules converges to the bootstrap minimization targets in Table \ref{alphaResultTable}. The coefficient $\eta$ converges well to the values fixed by unitarity (\ref{etaValsAlld}). In $9-d$, $\beta$ is clearly attaining values lower than the String Theory minimum. Similarly in $10-d$, the decreasing trend in $\beta$ appears to go below the string theory minimum. The $\beta$ values in $11-d$ are the noisiest and are far from converging.}
\label{fig:arcs11}
\end{figure*}

\section{Dispersive sum rules for $\alpha$, $\beta$ and $\gamma$} \la{ap:Arcs}

In this appendix, we obtain the low energy coefficients using fixed-$t$ dispersion relations for the 11d amplitude (in 9d and 10d the analysis is very similar).

Consider the function 
$f(s)\equiv \lim_{t\to0} \(A(s,t)-\frac{1}{stu}\).$
This function is analytic in the upper-half plane with a branch point at $s=0$. The branch cut stretches along the real axis and
$f(s^*)=\left[f(s)\right]^*$ on the first Riemann sheet. Furthermore, crossing symmetry implies that $f(s)=f(-s)$.
At small $s$, and for $d=11$, we can expand the function as follows,
\begin{equation}
f(s) = \alpha \,\ell_P^{6} + \eta \ell_P^{9} (s^{3/2} {+} (-s)^{3/2}) + 2\beta \ell_P^{10} s^2 + \mathcal{O}(s^{5/2}).
\end{equation}


Consider now the contour integral in figure \ref{fig:arcs} for the function $f(s)/s$.
By Cauchy's theorem $\oint ds\,f(s)/s=0$, which gives
\begin{align}
i \int_0^\pi d\theta\, f ( s_0 e^{i\theta}  )  &=\int_{s_0}^\infty \frac{ds}{s} \left[ f(s+i\epsilon) - f(-s+i\epsilon) \right] \nonumber \\
&=2 i \int_{s_0}^\infty \frac{ds}{s} {\rm Im}\, f(s+i\epsilon)  
\label{idCauchy}
\end{align}
For $s_0\ll1$, the left hand side of \eqref{idCauchy}, becomes
\beq
i \int_0^\pi d\theta\, f ( s_0 e^{i\theta}  ) =i\pi \alpha \ell_P^6 - i\frac{4\eta}{3\pi} \ell_P^9 s_0^{3/2} + \mathcal{O}(s_0^{5/2}).
\eeq
In the same limit, the right hand side of \eqref{idCauchy} gives
\beq
2i\int_{s_0}^\infty \frac{ds}{s}\, {\rm Im}\, f(s^+)\equiv 2i  \int_{s_0}^\infty \frac{ds}{s}\, {\rm Im}\, A(s^+,t=0) \, ,
\eeq
where $s^+=s+i\epsilon$. Therefore, we conclude that
\beq
 \alpha - \frac{4\eta}{3\pi} \(\ell_P^2 s_0\)^{3/2} \approx \frac{2}{\pi} \int_{s_0}^\infty ds\, \frac{ {\rm Im}\, A(s^+,0) }{\ell_P^6 s} \,.
\label{alphasumrule}
\eeq

Similarly, we obtain a sum rule involving $\beta$ by repeating this analysis with $f(s)/s^3$ 
\begin{equation}
    \beta + \frac{2\eta}{\pi} \(\ell_P^2s_0\)^{-1/2}\approx\frac1\pi \int_{s_0}^\infty d s\, \frac{ {\rm Im}\ A(s^+,0) }{ \ell_P^{10}s^{3}}\ .
\end{equation}
Choosing the function $f(s)/s^{5/2}$
yields the following sum rule,
\begin{align}
    \eta -\frac{2\alpha}{3\pi} &\(\ell_P^2s_0\)^{-3/2} + \frac{4}{\pi}\beta \(\ell_P^2s_0\)^{1/2} \approx \nonumber\\
    &\frac2\pi \int_{s_0}^\infty d s\, \frac{1}{\ell_P^9 s^{5/2}}\, {\rm Im}\[\frac{ A(s^+,0) }{(1+i)}\].
\end{align}

For completeness, we report the sum rules in 9d
\begin{align}
 &   \alpha+\frac{4\eta}{\pi}(\ell_P^2 s_0)^{1/2}\approx \frac{2}{\pi}\int_{s_0}^\infty ds \frac{{\rm Im}\, A(s^+,0)}{\ell_P^6 s}\\
&\eta+\frac{2\alpha}{\pi}(\ell_P^2 s_0 )^{-1/2}-\frac{4\beta}{3\pi}( \ell_P^2 s_0)^{3/2}\approx\nonumber \\ &\quad \quad \qquad \qquad \qquad \frac{2}{\pi}\int_{s_0}^\infty \frac{ds}{\ell_P^7 s^{3/2}} {\rm Im}\, \left[\frac{A(s^+,0)}{(1-i)}\right]\\
&\beta-\frac{2\eta}{3\pi}(\ell_P^2 s_0)^{-3/2}\approx \frac{1}{\pi}\int_{s_0}^\infty ds \frac{{\rm Im}\, A(s^+,0)}{\ell_P^{10} s^3}
\end{align}

As explained in the main text, we can use the sum rules in \eqref{10dSumRules} and those computed above to extract $\beta$ and the 1-loop coefficient $\eta$ from the extremal amplitudes. 
For instance, in 10d we evaluated the sum rules at various values of $N$ and $L$, with $s_0$ lying in the region depicted in figure \ref{fig:integrandAlpha} and found the best fit values for $\alpha, \beta$ and $\eta$ (see figure \ref{fig:arcs11}). In all dimensions, $\eta$ converges to the value predicted by unitarity. In 10d and 11d, $\alpha$ converges to values slightly below those in Table \ref{alphaResultTable} due to the following reason - the ansatz (\ref{ansatz0}) behaves like $s^{9/2}$ at low energies, whereas the expected low energy behaviour, from unitarizing the tree level is $s^{d/2}$. This mismatch for $d=10$ and $11$ leads to the peak at $s=0$ seen in figure \ref{fig:integrandAlpha}, and ignoring that leads to slightly lower $\alpha$.

\section{Resonances in the Extremal Amplitudes}
\label{ap:resonances}

\begin{figure}[t]
\centering
        \includegraphics[scale=0.30]{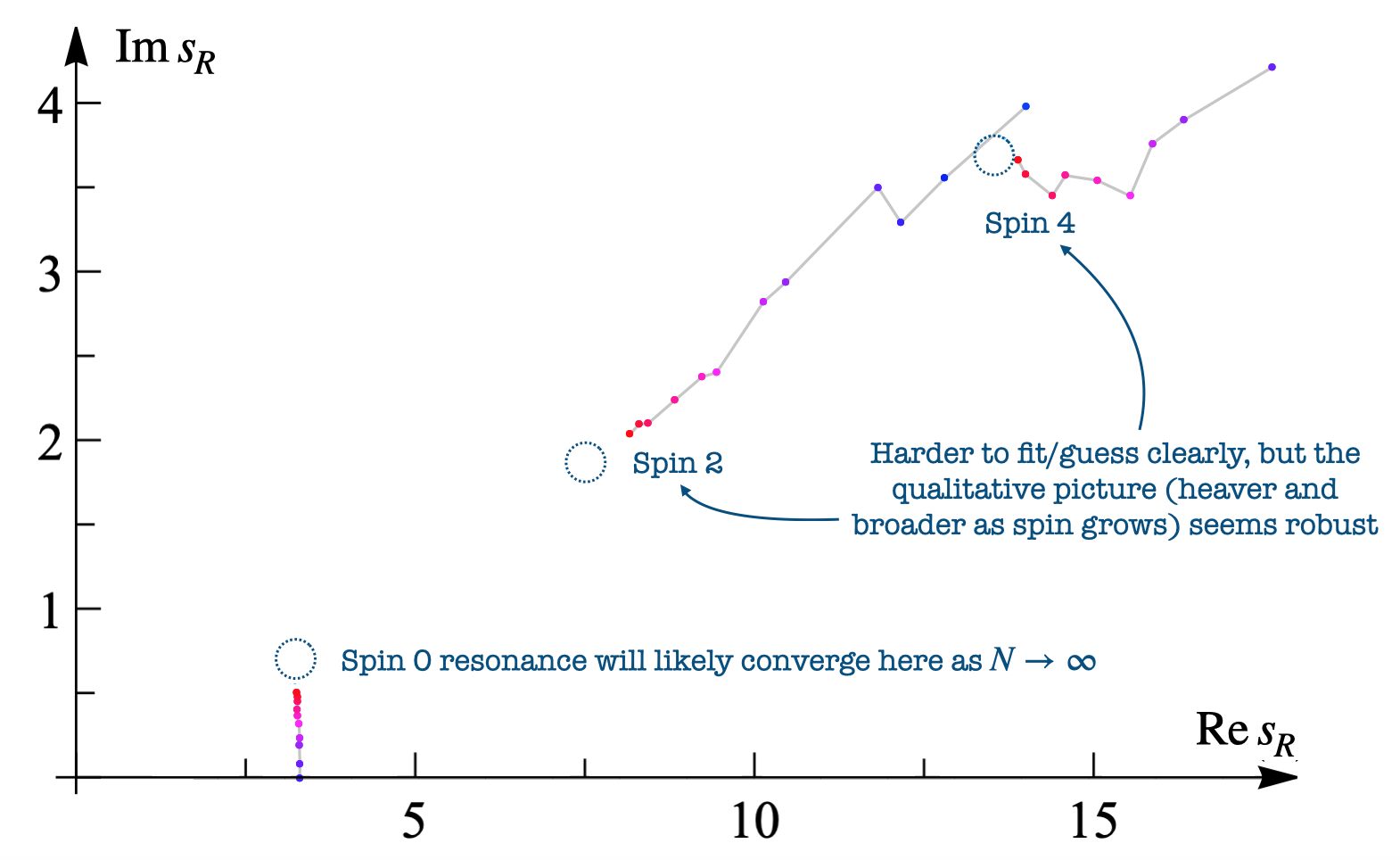}
\vspace{-0.5cm}
\caption{Lightest resonances' location $s_R$ for spin $l=0,2,4$ as we increase $N$ all the way to $N=30$. In a dashed blue circle we estimated (by eye) where the resonance seems to eventually converge to.}
\label{resonancesMoving}
\end{figure}

Resonances in a scattering amplitude show up as zeros of the partial waves $S_l(s_R)=0$ on the physical sheet of the complex $s$ plane. 
When the zero is sufficiently close to the real axis in a region where the scattering is nearly elastic $|S_l( {\rm Re}\, s_R)|\sim 1$ it produces the expected experimental signature of an unstable particle: the spin $l$ phase shift $\delta_l(s)=\tfrac{1}{2i}\log{S_l(s)}$ jumps by $\pi$ in a neighbourhood  of ${\rm Re}\, s_R$. In this regime 
we can use the Breit-Wigner approximation and relate the position of the zero to the physical mass and decay width parameters $s_R=(m_R+\tfrac{i}{2}\Gamma_R)^2$.
In figure \ref{resonancesMoving} we plot the location of the three most visible zeroes corresponding the lowest lying resonances in the spin $l=0,2,4$ for the extremal amplitude in $d=10$ as a function of $N$. \footnote{
To find the position of the resonances we perform a Newton's search in the complex plane. At each step we approximate the zero by 
\beq
s_{n+1}=s_n-\frac{S_l(s_n)}{S^\prime_l(s_n)}
\eeq
where the values of $S_l(s_n)$ and $S_l^\prime(s_n)$ are computed numerically using a simple quadrature method.
} The dependency on $N$ is denoted by the color (from blue to red as we increase $N$). It is interesting to notice how the ratio $\tfrac{\Gamma}{m}\sim 0.2$ for these three resonances, is comparable with the ratio $\tfrac{\Gamma_\rho}{m_\rho}$ for the $\rho$ meson in QCD. This confirms that the spectrum of resonances of the extremal amplitude minimizing $\alpha$ resembles that of a strongly coupled theory.
Figure \ref{resonancesMoving} also shows the different rate of convergence of the different partial waves. For the spin zero graviball, the location of the real part of the resonance almost does not change as we we increase $N$, although it moves along the imaginary axis entering in the physical sheet through the cut. For the higher spin resonances convergence is harder and their trajectory in $N$ looks more erratic. The blue circles denote our extrapolated estimates.

\begin{figure*}[t]
    \subfloat[\label{subfig:a}Spin-0 resonances]{\includegraphics[width=0.75\linewidth]{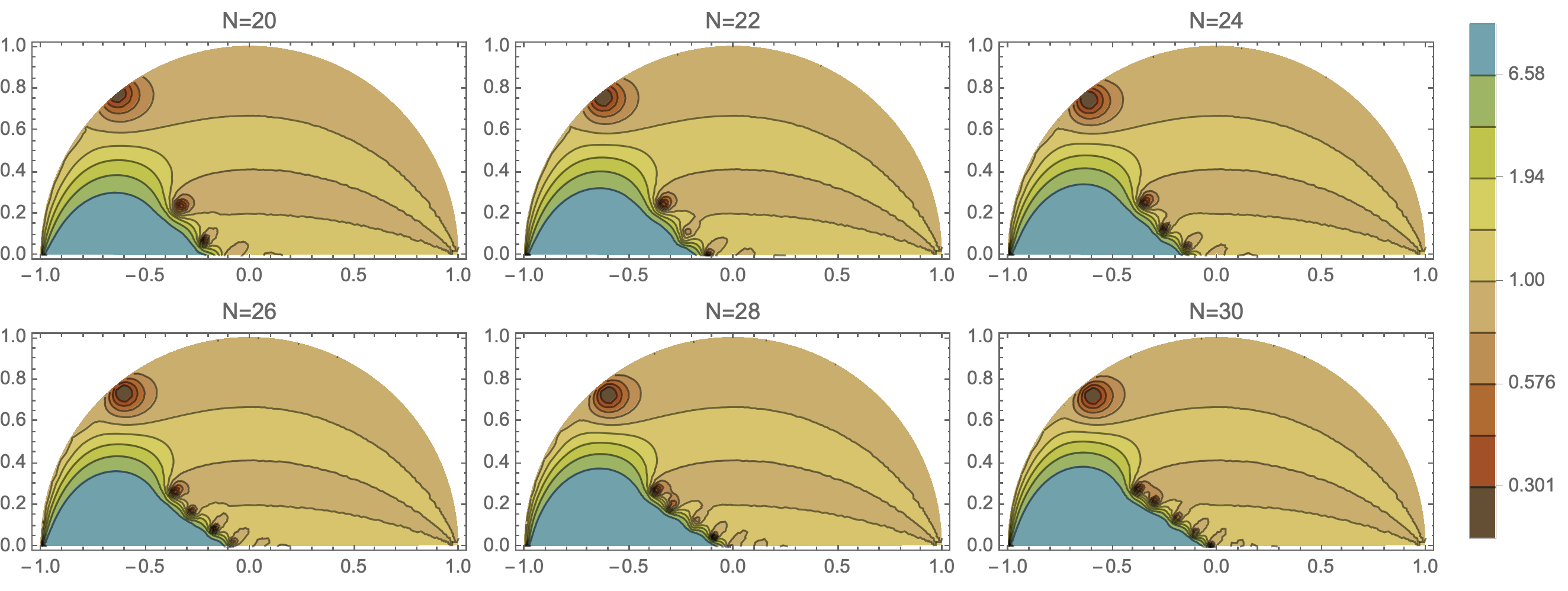}}\hfill\\
    
    \subfloat[\label{subfig:b}Spin-2 resonances]{\includegraphics[width=0.75\linewidth]{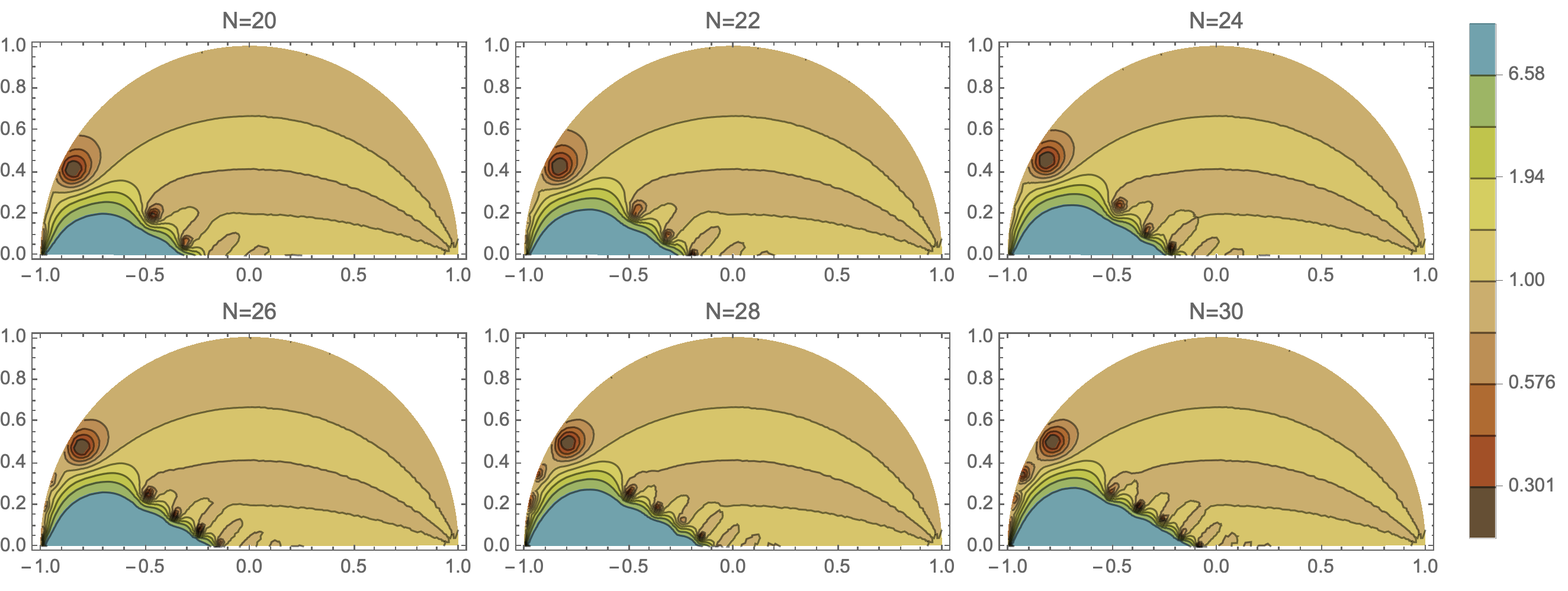}}\hfill\\
    
    \subfloat[\label{subfig:c}Spin-4 resonances]{\includegraphics[width=0.75\linewidth]{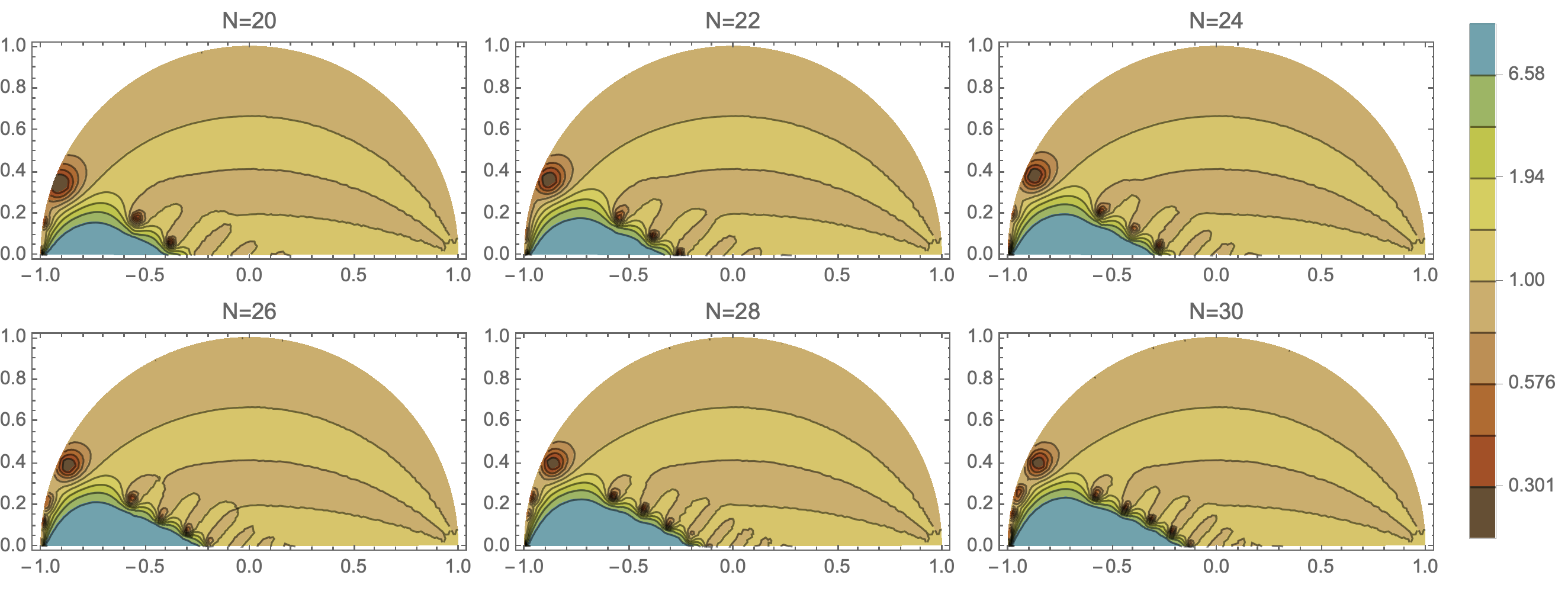}}\hfill\\
    \caption{Resonances in the 10d extremal amplitudes for various spins and $N$ in the $\rho$ plane. Here, the physical energies $s>0$ are mapped to the boundary of the disks and we see resonances entering into the first sheet as we increase $N$}
\end{figure*}

In figure \ref{subfig:a}, \ref{subfig:b}, and \ref{subfig:c} we plot the absolute value of $|S_l(\rho)|$ for $l=0,2,4$ in the upper half-plane in $s$, corresponding to the semi-circle $|\rho|\leq 1$, and ${\rm Im}\, \rho \geq 0$, for different values of $N$.
For the spin zero partial wave, figure  \ref{subfig:a}, we observe the presence of a single resonance close to the boundary of the disk (the spin zero graviball of figure \ref{resonancesMoving}), and a number of zeros entering through the left cut $-1<\rho<1$. Their number increases as we increase $N$. We do not have a physical interpretation for such zeros since they lie almost along the imaginary $s$ axis and cannot be interpreted as resonances. For spins $l=2,4$ in figures \ref{subfig:b}, and \ref{subfig:c} beyond the leading $l$-spin graviballs we also observe several higher energy resonances entering into the $\rho$ disk through the boundary, accompanied by other zeros coming in through the left cut. 

The way new resonances enter the physical sheet is quite interesting. On the one hand unitarity always tends to be saturated in these S-matrix bootstrap numerics \cite{4dpaper1}, trying to converge towards $|S_l(s)| \simeq 1$ for real $s$. (Indeed, note in figure \ref{spin0} that for $s>0$ and not too large we have indeed $|S| \simeq 1$ indicated by the orange colour; that region becomes larger as $N$ increases.) On the other hand, when the resonance zero enters the physical sheet we will have $|S_l(s)| \simeq 0$ nearby that resonance. In practice what happens is that unitarity needs to be sacrificed for a while when the resonance enters the physical sheet and afterwards it struggles to be saturated again. 
In other words, as we increase $N$, there is a tension between unitarity saturation and the addition of more and more resonances needed to ensure a good behavior of the amplitude in the Regge limit. This should be related to the tension raised by Caron-Huot which we discussed in appendix \ref{runawayappendix}. 

In figure \ref{fig:my_label} we plot the $|S_l(\rho)|$ for $l=0,2,4$ in 9d and 11d. The extremal amplitudes in these cases present the same qualitative features of the 10d amplitudes.

\begin{figure*}
    \includegraphics[width=0.75\linewidth]{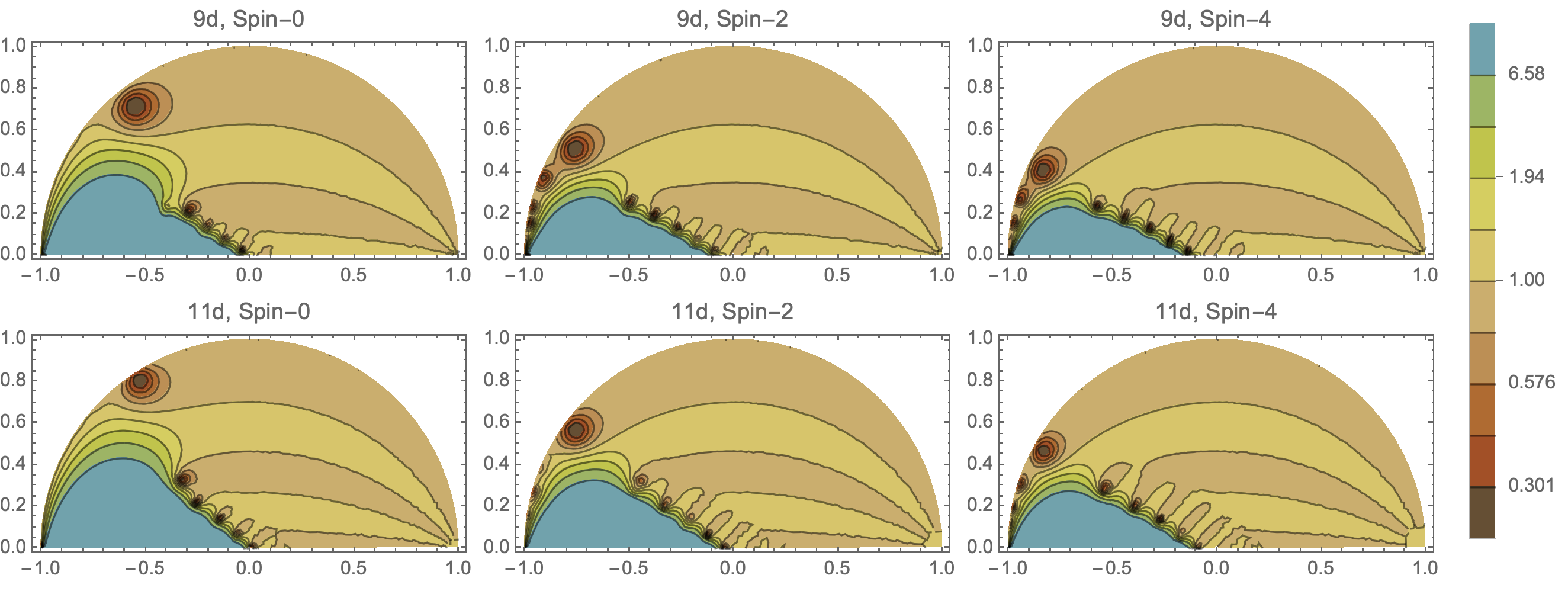}
    \caption{Resonances in 9d and 11d for $N=30$ and $L=200$. Here too, the spin-0 partial wave only has one resonance close to the real axis in the $s$ complex plane.}
    \label{fig:my_label}
\end{figure*}

\section{Virasoro Shapiro Spin Decomposition} \la{VSappendix}
It is amusing to recall how $\alpha$ extracted from the sum rule goes for perturbative string theory and contrast that to what we find here. By expanding Virasoro-Shapiro we have 
\beq
A_\text{VS}=\frac{1}{s t u} \frac{\Gamma(1-\frac{s}{4})\Gamma(1-\frac{t}{4})\Gamma(1-\frac{u}{4})}{\Gamma(1+\frac{s}{4})\Gamma(1+\frac{t}{4})\Gamma(1+\frac{u}{4})} \simeq \frac{1}{s t u} + \frac{\zeta_3}{32} +\dots \eeq
Here we omit the $8\pi G_N$ overall normalization and recall that $s,t,u$ are all measured in units of string length (and not Planck length). Let us recall first how the sum rule
\beq
 \alpha  = \frac{2}{\pi} \int_{0}^\infty ds\, \frac{ {\rm Im}\, A(s+i\epsilon,t=0) }{s  } \,. \label{aRule}
\eeq
produces the perturbative $\alpha_\text{VS}=\zeta_3/32$.
(Again, the $\zeta_3/32$ is in units of string length; in units of Plank Length it would be $\zeta_3/32 \times (l_\text{string}/l_\text{Plank})^6 \to \infty$ in this perturbative regime.)
When $t=0$ we have $u=-s$ and  
\beq
A_\text{VS} \simeq -\frac{1}{s^2 t}+\Big(\frac{1}{s^3}-\frac{1}{s^2} H(s/4)+\frac{1}{s^2} H(-s/4)\Big) +\dots
\eeq
At positive $s$ the last Harmonic number has poles when $s/4$ in an integer; these correspond to the various massive strings. At the pole $s=4n$ we thus have
\beq
A_\text{VS} \simeq -\frac{1}{(4n)^2 t}-\frac{1}{(4n)^2}\times \frac{1}{s-4n} +\dots
\eeq
Each pole contributes as $-i \pi$ to the imaginary part, 
\beq
\text{Im}(A_\text{VS}) \simeq \frac{\pi}{(4n)^2}\delta(s-4n) 
\eeq
so that plugging it into (\ref{aRule}) yields
\beq
 \alpha_\text{VS}  =2 \sum_{n=1}^\infty \frac{1}{(4n)^3} =\frac{\zeta_3}{32} \,.
 \label{aRule}
\eeq
It is fun to probe deeper inside this sum rule and decompose it into its various spin contributions. We have, for $s\simeq 4n$,
\beqa
A_\text{VS} &\simeq& 
\frac{1}{s-4n} \times \frac{(-1)^{n+1} \Gamma\left(-\frac{t}{4}\right) \Gamma \left(n{+}\frac{t}{4}{+}1\right)}{(4 n{+}t)^2 \Gamma (n{+}1)^2 \Gamma \left(\frac{t}{4}{+}1\right)\Gamma \left({-}n{-}\frac{t}{4}\right)} \nn \\
   &\equiv& \frac{1}{s-4n} \sum_{l=0}^{2(n-1)} c_l^{(n)} \frac{C_l^{7/2}(1+\tfrac{2t}{s})}{C_l^{7/2}(1)} \,. 
\eeqa
It is trivial to compute as many constants $c_l^{(n)}$ as needed; it is quite nontrivial to establish analytically that they are all negative as recently studied in \cite{Arkani-Hamed:2022gsa}, see also~\cite{Maity:2021obe}. 

The previous computation is reproduced as 
\beq
\alpha_\text{VS}  =\frac{2}{\pi} \sum_{n=1}^\infty \frac{-\pi}{4n} \underbrace{ \sum_{l=0}^{2(n-1)} c_{l}^{(n)}}_{-\frac{1}{(4n)^2}} \,.
\eeq
Instead, we can swap both sums to get 
\beq
\alpha_\text{VS}  = \sum_{l=0}^\infty \Big(\alpha_\text{VS}^{(l)} \equiv\sum_n \frac{-1}{2n} c_{l}^{(n)} \Big) \,.
\eeq
We can plot how the contribution of this sum up to spin $L$ contributes to $\alpha$. Spin $0$ alone contributes to about $84\%$ of the result and the first $4$ spins capture $99\%$ of $\zeta_3/32$ as depicted in figure \ref{fig:alphaDist}. We can of course repeat the same exercise and see how much the various spins contribute to our bootstrapped amplitude; that is also in figure \ref{fig:alphaDist} for $d=10$. ($d=9,11$ exhibit a very similar behavior.)

Once we find an interpolation between perturbative and strongly coupled String theory we should find an interpolation between all such sum rules. 

\bibliography{WhereIsMtheory} 

\end{document}